\documentclass[aps,prl,twocolumn
]{revtex4-2}
\pdfoutput=1
\usepackage[english]{babel}
\usepackage{color,graphicx}
\usepackage[usenames,dvipsnames]{xcolor}
\usepackage{amsmath,dsfont,url,amssymb}
\usepackage{slashed,cancel}
\usepackage[usenames,dvipsnames]{xcolor} 
\usepackage[colorlinks=true, linkcolor=black, citecolor=ForestGreen]{hyperref}
\usepackage{mdframed}
\usepackage{subfigure,overpic}
\usepackage{feynmf}
\usepackage[compat=1.1.0]{tikz-feynman}
\usepackage{textgreek}
\usepackage{ulem}

\newcommand{\wn}{{\mathfrak{w}}}

\newcommand{\so}[1]{{}}
\newcommand{\nw}[1]{#1}

\begin{document}

\title{Theory of non-linear diffusion with a physical gapped mode}

	\author{Navid Abbasi} \email{abbasi@lzu.edu.cn} \affiliation{
	School of Nuclear Science and Technology, Lanzhou University, 222 South Tianshui Road, Lanzhou, 730000, China}
\author{Matthias Kaminski} \email{mski@ua.edu} \affiliation{
	Department of Physics and Astronomy, 
	University of Alabama, 
	Tuscaloosa, AL 35487, USA}
\author{Omid Tavakol} \email{otavakol@physics.utoronto.ca} \affiliation{Department of Physics, University of Toronto, Toronto, Ontario M5S 1A7, Canada}

\date{\today}

\begin{abstract}
In a system with one conserved charge the charge diffusion is modified by non-linear self-interactions within an effective field theory (EFT) of diffusive fluctuations. 
We include the slowest ultraviolet~(UV) mode, constructing a UV-regulated EFT. 
The relaxation time of this UV mode is protected from renormalization, as supported by experimental data in a bad metal system. 
Furthermore, the retarded density-density Green's function acquires four branch points, eventually increasing the range of applicability. 
We discuss the fate of long-time-tails as well as implications for the quark gluon plasma.
\end{abstract}
\maketitle

\section{Introduction}
In a system with a single conservation law, the late-time dynamics is universally described by  hydrodynamic diffusion of the associated conserved charge. In the EFT approach to hydrodynamics~\cite{Crossley:2015evo,Jensen:2018hse,Haehl:2015pja,Glorioso:2018wxw}, the linear diffusion equation is derived as the equation of motion from a classical quadratic effective action. If the diffusion constant depends on the transported density, then this equation becomes non-linear~\footnote{This was discussed in \cite{Kovtun:2014nsa,Chen-Lin:2018kfl,Chao:2020kcf}. While \cite{Chen-Lin:2018kfl} focuses on density fluctuations in the framework of Schwinger-Keldysh, \cite{Chao:2020kcf} studies multiplicative noise, arisen from such non-linear dependence in the context of stochastic fluid dynamics. See also~\cite{Chao:2023kvz}.}.  
This is caused by the (self-)interactions encoded in the corresponding effective action, eventually affecting the correlation functions~\cite{Chen-Lin:2018kfl,Chao:2020kcf}.

The onset of diffusion in a system can be characterized by the \textit{relaxation time} $\tau$, the timescale of the relaxation of that non-conserved quantity which relaxes slowest. 
In the limiting case $\tau=0$, the interacting EFT of hydrodynamics predicts long-time-tails~\cite{Kovtun:2003vj} and large renormalization effects for the transport coefficients \cite{Kovtun:2011np}. In this limit, which corresponds to the instantaneous response of the current to the source, hydrodynamic equations fail to satisfy the commutation sum rules~\cite{kadanoff1963hydrodynamic}. 
Therefore a finite relaxation rate $1/\tau$ has to be considered to regulate the theory of diffusion   discussed above. Such a relaxation rate has been recently measured in  
ultra-cold gas of  $^6\text{Li}$ in a two-dimensional lattice, an example of a ``bad metal'', as the testing ground for the Fermi-Hubbard model \cite{Brown:2018}.

The aforementioned relaxation time $\tau$ characterizes the timescale of the establishment of local thermal equilibrium in a U(1) static system~\cite{Hartnoll:2021ydi}~\footnote{When the system is not static, flow dependent timescales are also involved in thermalization. Examples are the weakly coupled system of~\cite{Behtash:2019txb} under the Bjorken flow, and~\cite{Bazow:2015dha} under the FLRW expansion.}. 
Since the local equilibration cannot happen arbitrarily quickly, it is expected for $\tau$ to be subject to a lower bound~\cite{Hartnoll:2021ydi}.
It is then important to understand how this bound is affected by fluctuations.   
This can be investigated via studying the effect of self-interactions in the EFT framework, in particular the renormalization effects. In fact, we need to increase the EFT cutoff of \cite{Chen-Lin:2018kfl,Chao:2020kcf} to include at least the slowest UV mode. 
Indeed, this leads us to derive, for the first time, a UV-regulated \textit{theory of non-linear diffusion with a physical gapped mode}.

Another motivation manifests itself in the long-time-tails. It is well known that in a many body system, at late times, the auto-correlation functions feature long-time-tails~\cite{ernst1970asymptotic,alder1970decay}. 
Now the question is what happens to these tails when the local thermalization is not instantaneous. Does the thermalization occur exponentially fast~\footnote{We focus on static systems. Systems in a dynamical background may not exhibit such exponential behavior; for example the Gubser flow~\cite{Gubser:2010ze}.}? 
If not, would it possibly lead to a longer tail for correlation functions, thereby delaying the global thermalization?

Keeping these questions in mind, in the following, we construct the EFT including fluctuations for a self-interacting system with a single conserved density together with a single gapped mode; the longest-lived gapped mode, acting as a UV-regulator for the EFT. 
Our effective action is consistent with the general framework of \cite{Crossley:2015evo}, however, derived through the  traditional  Martin-Siggia-Rose (MSR) formalism \cite{martin1973statistical}. Explicitly calculating the one-loop retarded density-density Green's function, we discuss how the renormalization of the diffusion constant at $\tau=0$, performed in \cite{Chen-Lin:2018kfl,Chao:2020kcf}, is affected by the gapped mode. In particular we discuss the renormalization of the relaxation time which was not included in \cite{Chen-Lin:2018kfl,Chao:2020kcf}. The renormalized Green's functions then will be used to study the effect of the gapped mode on long-time-tails.  
We conclude by discussing application of our results to a bad metal~\cite{Brown:2018}, expanding quark gluon plasma (QGP) and also to the QGP droplet near the QCD critical point. 
Natural units, $k_B=1=\hbar$, are implied. 

\section{Setup}
Beginning in the absence of interactions, we consider a $U(1)$ conserved charge with a non-relativistic non-conserved current as follows
\begin{equation}\label{dynamic_eq}
	\partial_t n +\boldsymbol{\nabla}\cdot \textbf{J}=\,0\,,\,\,\,\,\,\tau\, \partial_t \textbf{J}+\textbf{J}+ D \,\boldsymbol{\nabla}n=\,0 \, ,
\end{equation}
with $D$ being a constant. 
For finite $\tau$, the combination of the two equations in~\eqref{dynamic_eq} leads to 
\begin{equation}\label{MIS}
	\tau \partial_t^2  n +\,\partial_t  n-\,D \boldsymbol{\nabla}^2  n=\,0 \, .
\end{equation}
%
This equation describes the dynamics of the diffusive density mode together with a gapped mode 
\begin{equation}\label{mode_no_int}
	\omega_{1,2}=\,-\frac{i}{2 \tau}\big(1\mp \sqrt{1 -4 \tau D \textbf{k}^2}\big) \, .
\end{equation}
The theory has two timescales: $\tau_{UV}\equiv - i /\omega_1$ is the timescale of the thermalization while $\tau_D\equiv - i /\omega_2$ is the diffusion time of the mode with momentum $\textbf{k}$. 
For the long wavelength modes with $\tau\ll (D\textbf{k}^2)^{-1}$:
\begin{equation}\label{}
	\omega_{1}=- i D \textbf{k}^2,\,\,\,\,\,\,\omega_{2}=-\frac{i}{\tau}+i  D \textbf{k}^2 \, ,
\end{equation}
which gives $\tau_{D}=(D\textbf{k}^2)^{-1}$ and $\tau_{UV}=\tau$.
This can be compered to Fick's law of diffusion, $\partial_t  n-\,D \boldsymbol{\nabla}^2  n=\,0$, which encodes one single diffusion mode $\omega=- i D \textbf{k}^2$. Since $\omega_2$ is a UV mode for Fick's law, we refer to \eqref{MIS} as the (phenomenological) 
\textit{theory of diffusion, UV-regulated with a physical gapped mode}. 

Our intent is to make equation \eqref{MIS} non-linear. When $n$ does not couple to any dynamical field, this is achieved by including the self-interactions of $n$ \cite{Kovtun:2014nsa}. These enter by  promoting $D$ to be a function of small fluctuations about $n=0$:
$\,D(n)=\,D+\,\lambda_D\, n+\,\frac{\lambda'_D}{2}\, n ^2+\cdots$, where $\lambda_D=dD(n)/dn, \lambda_D'=d^2D(n)/dn^2, \cdots$.
Dropping terms with orders higher than $2$, the non-linear version of \eqref{MIS} is found to be
\begin{equation}\label{nonlinear_eq}
	\text{E}[n]\equiv\, \tau \partial_t^2 n +\,\partial_t  n-\, \boldsymbol{\nabla}^2 \big(D\, n+\frac{\lambda_D}{2} n^2+\frac{\lambda'_D}{6} n^3\big)=\,0 \, .
\end{equation}
 In general $\tau$ could  be a function of $\textbf{J}$ in \eqref{dynamic_eq}. However, since there is no other vector involved in the setup, the scalar $\tau$ cannot depend on the vector $\textbf{J}$ at linear order; the first contribution is quadratic in $\textbf{J}$ and turns out not to contribute to any of our results. We take $\tau$ constant in what follows. Note also that if $n$ refers to a charge density, then under its changing sign, $\lambda_D\rightarrow -\lambda_D$ but $D$ and $\lambda'_{D}$ remain the same.

Our goal is to study correlations between fluctuations of $n$. For this, we construct an effective action for $n$ whose equation of motion is \eqref{nonlinear_eq}.
We do this in the framework of  MSR formalism~\cite{martin1973statistical} . The idea is to put a noise term on the right side of \eqref{nonlinear_eq} and then impose the fluctuation-dissipation theorem to fix its strength. Finally,  exponentiating this stochastic equation yields the effective action $S_{\text{eff}}=\int dtd^dx \mathcal{L}_{\text{}}$. As we show in the Supplemental Material, 
\nw{
the effective action takes the form
\begin{equation}\label{}
\mathcal{L}=iT \sigma(n)(\boldsymbol{\nabla}n_a)^2-n_a\left(\tau \partial_t^2n+\partial_t n- \boldsymbol{\nabla}D(n) \boldsymbol{\nabla}n\right)+\cdots
\end{equation}
where the charge conductivity $\,\sigma(n)=\sigma+\lambda_{\sigma} n+\frac{\lambda'_{\sigma}}{2} n ^2+\cdots$, is related to the diffusion coefficient by the Einstein relation $\sigma =\chi D$ via the charge susceptibility $\chi$.  
To quartic order in fields it becomes } 
\begin{equation}\label{quad_eff_L}
	\begin{split}
		\mathcal{L}&=iT \sigma(\boldsymbol{\nabla}n_a)^2-\, n_a\left(\tau \partial_t^2n+\partial_t n-D \boldsymbol{\nabla}^2 n\right)\\
		&+\,iT\chi \lambda_{\sigma}\,n(\boldsymbol{\nabla}n_a)^2+\frac{\lambda_D}{2}\, \boldsymbol{\nabla}^2	n_a\, n^2 \\
		&+\frac{1}{2}iT\chi \lambda'_{\sigma}\,n^2(\boldsymbol{\nabla}n_a)^2+\frac{\lambda'_D}{6}\, \boldsymbol{\nabla}^2	n_a\, n^3 \, . 
	\end{split}
\end{equation}
\so{where $\,\sigma(n)=\,\sigma+\,\lambda_{\sigma}\, n+\,\frac{\lambda'_{\sigma}}{2}\, n ^2+\cdots$ and $\chi$ is the susceptibility.}
Here $n_a$ is an auxiliary field, analogous to the $a$-field in the Schwinger-Keldysh framework~\cite{Chen-Lin:2018kfl}.    
In the limit $\tau =0$, \eqref{quad_eff_L}  reduces to the  theory of \textit{diffusive fluctuations}~\cite{Chen-Lin:2018kfl}. 
At $\tau\neq 0$, \eqref{quad_eff_L} defines our novel {\it theory of non-linear diffusion with a physical gapped mode}.

In order to study the leading effects caused by the non-linear terms in \eqref{nonlinear_eq},  we investigate the effect of one-loop corrections on the retarded Green's function in the EFT described by \eqref{quad_eff_L}. At one-loop order, only cubic interactions in \eqref{quad_eff_L} contribute.  This is why we dropped higher order terms when expanding $D(n)$ and $\sigma(n)$. Even $\lambda_D'$ and $\lambda_{\sigma}'$ are not needed. 

\section{Results}
The retarded Green's function at one-loop can be parameterized by 
\begin{equation}\label{G_R}
	G^{R}_{nn}(\omega,\textbf{k})=\,\frac{i\,\big(\sigma+\delta \sigma(\omega,\textbf{k})\big)\,\textbf{k}^2}{- i \tau \omega^2+ \omega +i D \textbf{k}^2+ \Sigma(\omega,\textbf{k}) }\,.
\end{equation}
The self-energy is given by
\begin{equation}\label{Sigma_00}
	\begin{split}
		&	G^{(0)}_{n n_a}\Sigma\so{(p)}G^{(0)}_{n n_a}=\\\,
		&	\scalebox{0.45}{	\begin{tikzpicture}[baseline=(a.base)]
				\begin{feynman}
					\vertex (a) ;
					\vertex [right=of a] (a1) ;
					\vertex [right=of a1] (a2);
					\vertex [above right=of a2] (a3);
					\vertex [below right=of a2] (a4);
					\vertex [above right=of a4] (a5); 
					\vertex [ right=of a5] (a6);
					\vertex [ right=of a6] (a7);
					\diagram* {
						(a) --[very thick](a1)-- [boson,very thick] (a2) -- [quarter left, very thick] (a3)--[boson,quarter left,very thick](a5)--[quarter left, very thick](a4)--[quarter left,very  thick](a2) ,
						(a5) -- [very thick] (a6)--  [boson,very thick](a7),
					};
				\end{feynman}
		\end{tikzpicture}}
		+
		\scalebox{0.45}{	\begin{tikzpicture}[baseline=(a.base)]
				\begin{feynman}
					\vertex (a) ;
					\vertex [right=of a] (a1) ;
					\vertex [right=of a1] (a2);
					\vertex [above right=of a2] (a3);
					\vertex [below right=of a2] (a4);
					\vertex [above right=of a4] (a5); 
					\vertex [ right=of a5] (a6);
					\vertex [ right=of a6] (a7);
					\diagram* {
						(a) --[very thick](a1)-- [boson,very thick] (a2) -- [quarter left, very thick] (a3)--[quarter left,boson,very thick](a5)--[quarter left,boson,very thick](a4)--[quarter left, very thick](a2) ,
						(a5) --  [very thick](a6)--  [boson,very thick](a7),
					};\,.
				\end{feynman}
		\end{tikzpicture}} \, .
	\end{split}
\end{equation}
\begin{figure}[htbp] 
	\centering
	\centering
	\includegraphics[width=0.49\textwidth]{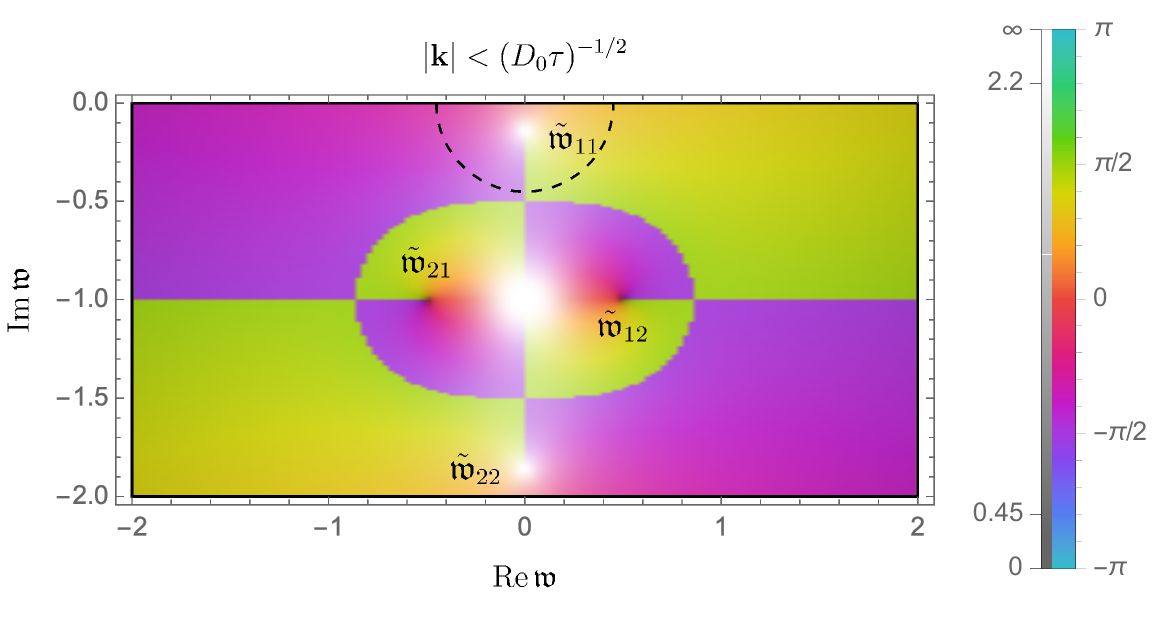} \\
	\caption{Contour plot of the phase angle, $\varphi$, of the complex-valued $\alpha_1=|\alpha_1| e^{i\varphi}$. 
		Any discontinuous transition between two distinct colors represents a branch-cut of the retreaded Green's function in the complex $\wn=\omega\,\tau$ plane. 
		For comparison, we have schematically illustrated the range accessible to the diffusive theory of Ref.~\cite{Chen-Lin:2018kfl} by a dashed semi-circle in the top panel, while our
		\textit{theory of non-linear diffusion (UV-regulated) with a physical gapped mode}  
		is valid in the entire range displayed.
	}
	\label{branch_cut}
\end{figure}
where from the first line in \eqref{quad_eff_L} we have $G_{n n_a}^{(0)}\so{(p)}=(\omega+ i D \textbf{k}^2- i \tau\, \omega^2)^{-1}$ \so{with $p\equiv(\omega, \textbf{k})$}.
Considering a hard momentum cutoff, we evaluate the loop integrals. 
The cutoff-dependent parts of $\Sigma$ can be absorbed into the bare values of coefficients $D$ and $\tau$. The cutoff-independent parts turn out to be non-analytic, that are given in $d=1,2,3$ dimensions by
\begin{equation}\label{sigma}
	\Sigma_d(\omega,\textbf{k})=\alpha_d(\omega,\textbf{k})(\tau D)^{\frac{2-d}{2}}\frac{ T \chi}{D^{2}} 
	\textbf{k}^2\bigg[f_{1d} \lambda_{D}^2+ 
	f_{2d}\lambda_{D}\lambda_{\sigma}\bigg] 
\end{equation}
Here and below $f_{i d}$ are analytic functions of \nw{$\omega$ and $\textbf{k}$}. The non-analyticity is encoded in 
\begin{eqnarray}\label{eq_alpha}
	\alpha_d(\omega,\textbf{k})=\frac{\big(- F(\omega, \textbf{k})\big)^{\frac{d}{2}-1}}{(16\pi)^{d/2}\Gamma(\frac{d}{2})}\cdot	\left\{
	\begin{array}{lr}
		i \pi& d \,\,\text{odd}\,\,\\
		\log\frac{1}{F(\omega, \textbf{k})} & \,\,d \,\,\text{even}  \\			
	\end{array}
	\right.
\end{eqnarray}
%
with $F(\omega, \textbf{k})=\frac{ (1- i \tau \omega)^2(D \textbf{k}^2\tau-i\omega\tau  (2-i\tau  \omega))}{D \textbf{k}^2 \tau +(1-i\tau  \omega)^2}$. 
Having found $\Sigma$, the next step is to calculate the loop correction to the numerator of $G^{R,(1)}_{nn}$. 	Similarly, we  find (see the Supplemental material)
\begin{equation}\label{delta_kappa_d}
	\frac{\delta \sigma_d(\omega,\textbf{k})}{\sigma}=\alpha_d(\omega,\textbf{k}) (\tau D)^{\frac{2-d}{2}}\,\frac{T \chi}{D^{2}}\textbf{k}^2\,\bigg[f_{3d}\lambda_D^2 + 
	f_{4d}\lambda_{D}\lambda_{\sigma}\bigg] \, .
\end{equation}
The two expressions \eqref{sigma} and \eqref{delta_kappa_d} fully specify \eqref{G_R}.
Rewriting it in the form 
\begin{equation}\label{G_R_c}
	G^{R}_{nn}(\omega,\textbf{k})=\,\frac{i\,\sigma\,\textbf{k}^2}{- i (\tau+\delta \tau(\omega,\textbf{k})) \omega^2+ \omega +i (D+\delta D(\omega,\textbf{k})) \textbf{k}^2}
\end{equation}
we find that
\begin{equation}\label{delta_D}
	\begin{split}
		\delta D(\omega,\textbf{k})&=\frac{\lambda_D^2 T \chi}{4D^{2}}(- i \omega)(\tau D)^{\frac{2-d}{2}}\alpha_d(\omega, \textbf{k})\\
		\times&(1-i\tau \omega)\left(\frac{2+D \textbf{k}^2\tau-\tau \omega(3i+\tau \omega)}{- D \textbf{k}^2\tau+(i +\tau \omega)^2}\right)^2 \, ,
	\end{split}
\end{equation}
and $\delta \tau(\omega, \textbf{k})=0$.
The latter answers the first question raised in the Introduction: 
\textit{our theory}, Eq.~\eqref{quad_eff_L}, predicts that the bound on local thermalization time $\tau$ (and thus the onset of diffusion)~\cite{Hartnoll:2021ydi} is protected from renormalization caused by the fluctuations. In the Discussion, we show that this result is supported by the experimental data associated with the bad metallic system of \cite{Brown:2018}.  Note that at $\tau=0$, \eqref{delta_D} reduces to the results found in \cite{Chen-Lin:2018kfl,Chao:2020kcf}.

Causality implies that $G^R$ is analytic in the upper half of the complex frequency plane. However, the structure of $F(\omega, \textbf{k})$ indicates that, due to the interactions,  four branch point singularities are induced in the lower half plane (Fig.~\ref{branch_cut}):
\begin{equation}\label{branch_points}
	\tilde{\omega}_{11,22}=-\frac{i}{\tau}\big(1\mp\sqrt{1-D \textbf{k}^2\tau}\big)\,, \quad
	\tilde{\omega}_{12,21}=-\frac{i}{\tau}\pm|\textbf{k}|\sqrt{\frac{D}{\tau}} \, .
\end{equation}
%
These branch points correspond to the minimum energy for \nw{generating two on-shell excitations with the dispersion \eqref{mode_no_int} in the loops}, as explained in the Supplemental Material. 
The location of branch points will be important when finding the inverse Fourier transform of momentum space correlation functions.

As the last formal result, we find $\mathfrak{G}_{nn}(t,\textbf{k})=\int \frac{d\omega}{2\pi}\,G_{nn}(\omega,\textbf{k})\,e^{-i \omega t}$. Note that $G_{nn}(\omega,\textbf{k})$ can be simply calculated from  $G^R_{nn}(\omega,\textbf{k})$ via the fluctuation-dissipation theorem. While at $\tau=0$ one finds $\mathfrak{G}_{nn}(t,\nw{\textbf{k}})$ analytically \cite{Michailidis:2023mkd}, at $\tau\ne0$ we were not able to represent  $\mathfrak{G}_{nn}(t,\textbf{k})$ in terms of known special functions. 
\so{This is because for the  Fourier integral one needs to deform the contour of integration in the lower complex frequency plane and pick up the discontinuities along the branch cuts. Fig.~\ref{branch_cut} shows that the branch cut structure at $\tau=0$ is a simple cut surrounded by a dashed semicircle. For $\tau\ne0$ however,  the existence of the other three branch points is an obstacle to expressing Fourier results with known functions.} 
Thus, we find this function numerically.

\begin{figure}[tb]
	\centering
	\centering
	\includegraphics[width=0.4\textwidth]{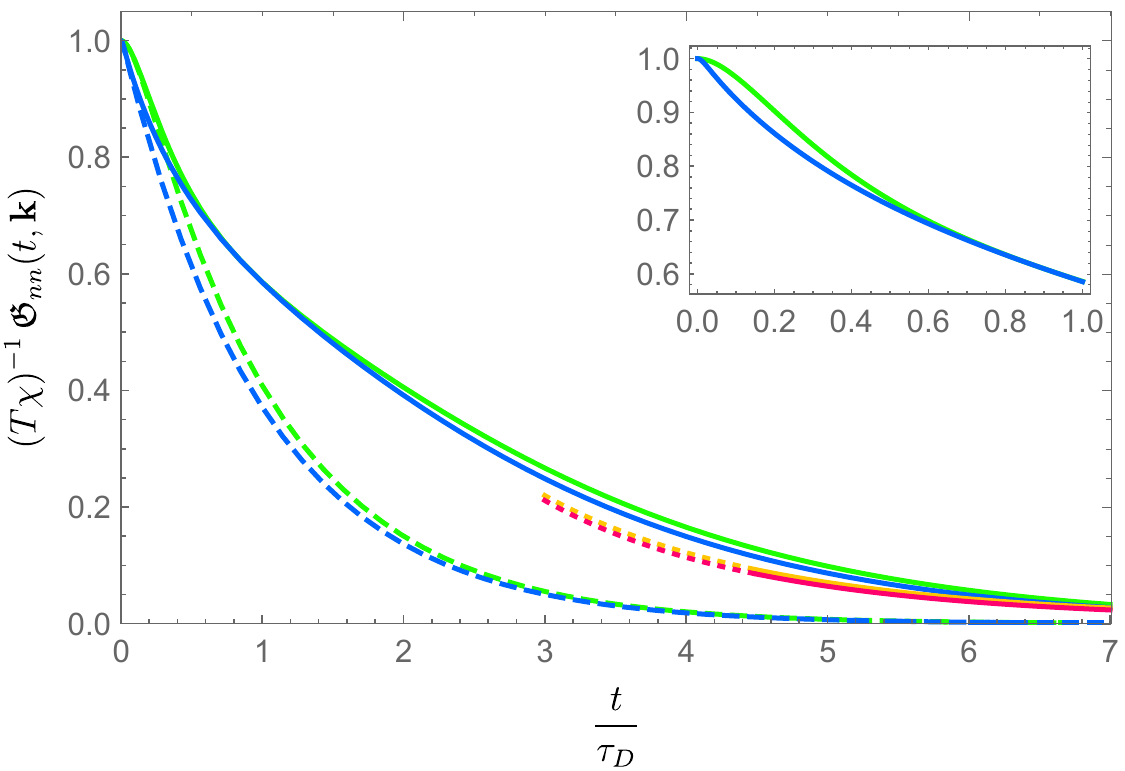}
	\caption{Solid blue and green curves show $\mathfrak{G}_{nn}^{(0)+(1)}$ while dashed curves correspond to $\mathfrak{G}_{nn}^{(0)}$. With blue and green, we illustrate {distinct} values of $r=\tau_{UV}/\tau_{D}$: $1/100 \,\text{and}\, 1/10$, respectively. The red and yellow curves show the long-time-tail of $\mathfrak{G}^{(0)+(1)}_{nn}$ for these two values of $r$. }
	\label{long_time_numeric}
\end{figure}

The results  \nw{for $\lambda_{\text{eff}}^2\equiv\frac{T\chi}{(\tau D^5)^{1/2}}\lambda_D^2=\frac{1}{2}$ } 
are displayed in Fig.~\ref{long_time_numeric}. 
For $t\gg\tau_{D}$, $r=\tau_{UV}/\tau_{D}$ has no significant effect; the two 
\nw{solid line } 
curves converge similarly. However, the effect of $r$ is significant at $t\ll\tau_{D}$. As shown in the inset, for $r=\frac{1}{10}$, the correlation function around $t=\frac{1}{10}\tau_{D}$ decays slower than for $r=\frac{1 }{100}$. This is related to the second branch point in Fig.~\ref{branch_cut}, i.e., $\tilde{w}_{22}$. As mentioned earlier, we did not find $\mathfrak{G}_{nn}$ analytically; but one would expect that when evaluating the Fourier integral, among other contributions, picking up the pole $\tilde{w}_{22}$ would yield a term $\propto e^{- i \tilde{\omega}_{22}t}\sim e^ {- t/\tau_{UV}}$. This would give a finite contribution at $t\lesssim  \tau_{UV}\sim r \tau_D$, which justifies the excess shown in the inset. Obviously, when $\tau_{UV}\rightarrow 0$  \cite{Chen-Lin:2018kfl,Chao:2020kcf}, $\tilde{w}_{22}$ becomes $-i\infty$, so the effect disappears. We emphasize that this feature is valid in static systems. In a dynamical background, the perturbations may not decay solely by an exponential term \cite{Heller:2015dha}.
 
 In the limit $\tau_D\ll t$, we find the asymptotic behavior of $\mathfrak{G}^{(1)}_{nn}(t,\textbf{k})$ analytically. This  ``long-time-tail", illustrated in red ($r=\frac{1}{10}$) and yellow ($r=\frac{1}{100}$)  in Fig.~\ref{long_time_numeric}, comes from integration over the region surrounded by the dashed semicircle in Fig.~\ref{branch_cut}. For $t>0$ and $d=1$:
\begin{equation}\label{longtime_tail}
	\mathfrak{G}^{(1)}_{nn}(t,\textbf{k})=g\frac{\big(1+\sqrt{1- \tau D \textbf{k}^2}\big)^4}{(1- \tau D \textbf{k}^2)^{1/4}}\frac{e^{(-1+\sqrt{1- \tau D \textbf{k}^2})\frac{t}{\tau}}}{\sqrt{2\pi D t}} \, ,
\end{equation} 
where $g=\frac{\lambda_{D}^2}{16D^2}T^2 \chi^2$.
This answers the second question raised in the Introduction: \textit{the fast thermalization of the non-conserved current} $\textbf{J}$ does not affect the fractional power of the long-time-tail. We find $t^{-1/2}$, which is identical to the known behavior $t^{-d/2}$ without UV regulator \cite{Kovtun:2003vj}. 
But the exponential decay associated with diffusive fluctuations changes. When $\tau=0$, the late-time behavior is given by $e^{-\frac{1}{2}D\textbf{k}^2t}$ \cite{Delacretaz:2020nit}. The effect of the gapped mode is to decrease this factor at any $\textbf{k}$. 
We emphasize that this exponential decay is specific to static systems near equilibrium. See \cite{Heller:2015dha} for a discussion on the existence of late-time universal attractors in a longitudinally expanding QGP. 

\section{Applications and Discussion} 
\begin{figure}[tb]
	\centering
	\centering
	\includegraphics[width=0.4\textwidth]{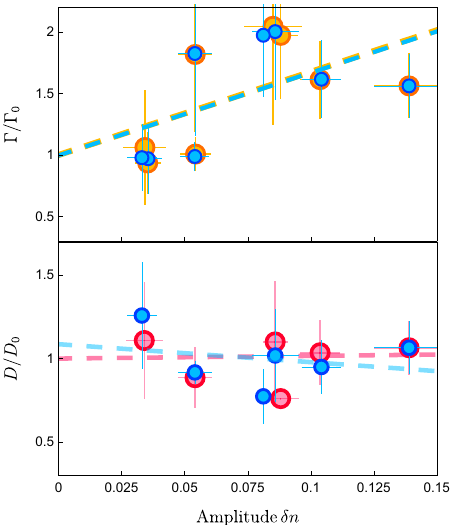}
	\caption{Variation of fitting parameters $\Gamma$ (top),  $D$ (bottom) versus the amplitude of the solution, namely $n(0,\textbf{k})$. 
	Orange and red dots are obtained by fitting the data of~\cite{Brown:2018} to Eq.~\eqref{MIS} while blue dots stem from fitting the data to Eq.~\eqref{nonlinear_eq}. 
	The orange, red, and blue dashed lines are the corresponding linear fits. 
	$\Gamma$ and $D$ are normalized by the extrapolated zero-amplitude values $\Gamma_0$ and $D_0$, respectively. \nw{Error bars indicate the standard error of the mean value resulting from our fits.}}
	\label{Fit}
\end{figure}

\textit{Bad metal: } 
\nw{
Let us check the results of our theory against data from the bad metallic system~\cite{Brown:2018} mentioned in the Introduction. 
In~\cite{Brown:2018}, the charge density $n(t, \textbf{k})$ is identified with twice the experimentally measured atomic density of one spin component. The data from~\cite{Brown:2018} consists of eight groups of points, each group of points contains ten $n(t, \textbf{k})$ values, with a fixed value of $\textbf{k} $. Ref.~\cite{Brown:2018} fits the data using the \textit{analytical} solution of \eqref{MIS} to determine $\Gamma$, $D$ and $\delta n\equiv n(0,\textbf{k})$. 
However, at high temperatures, which is the case in~\cite{Brown:2018}, the thermalization length becomes short, on the order of the lattice spacing, and fluctuation corrections are of order one \cite{Chen-Lin:2018kfl}. 
Therefore, instead of fitting data to the linear equation~\eqref{MIS}, as done in~\cite{Brown:2018}, we propose the non-linear equation~\eqref{nonlinear_eq}. 
Fig.~\ref{Fit} illustrates the results of this fit (blue) compared to the results of the linear equation fit (orange and red). For the sake of consistency, in both cases we fit data with the \textit{numerical} solution of the equation. For the linear case we find three parameters as in~\cite{Brown:2018}, while in the nonlinear case we also find a fourth parameter: $\lambda_D$. Our non-linear fit procedure and the dependence of our results on it are detailed step by step in the Supplemental Material.
}

\so{
In the bad metallic system~\cite{Brown:2018} mentioned in the Introduction, at high temperatures, the thermalization length becomes short, on the order of the lattice spacing, and fluctuation corrections to the conductivity are of order one \cite{Chen-Lin:2018kfl}. Therefore, instead of fitting data with a linear regulated diffusion equation~\eqref{MIS}, as done in~\cite{Brown:2018}, 
we propose our diffusion equation~\eqref{nonlinear_eq}. 
 Fig.~\ref{Fit} illustrates results of this fit (blue) compared to the results of the linear equation fit (orange and red); see Supplemental Material.
 }

As shown in the top panel, the linear fit of blue dots (blue line) is coincident with the linear fit of orange dots (orange line). 
Since each pair of orange-blue dots is associated with a specific $\textbf{k}$, this coincidence implies $\delta \Gamma(\textbf{k})\approx0$. This may be regarded as the realization of our earlier statement that $\tau=\Gamma^{-1}$ is protected from the non-linear corrections in our model. 

The bottom panel of Fig.~\ref{Fit} indicates that $\delta D(\textbf{k})$ is non-zero  for almost all values of $\textbf{k}$ in the experiment of~\cite{Brown:2018}. In other words, the diffusion constant is significantly renormalized by non-linear effects. This is in agreement with the prediction of our theory. 

\so{As pointed out in~\cite{Brown:2018}, their fit of the data to find $\Gamma$ and $D$ as functions of  initial amplitude is merely an attempt to evaluate the possibility of non-linear effects.    
In our work, we now see that the experimental data associated with the large initial amplitudes~\cite{Brown:2018} is qualitatively consistent with our theory. Additionally, the value of $\lambda_D=d D(n)/dn$ can be extracted from the data. 
} 
\nw{As pointed out in~\cite{Brown:2018}, their fit of large-initial-amplitude data with the linear equation is merely an attempt to evaluate the possibility of non-linear effects. We now see that the data is qualitatively consistent with our non-linear equation. Due to the large fluctuations existing in the system~\cite{Chen-Lin:2018kfl}, non-linearities must be taken into account.
Therefore our non-linear fit is preferred to~\cite{Brown:2018}'s fit as our fit has at least two quantitative advantages: First, the value of $D$ obtained here is more reliable than that of~\cite{Brown:2018}. Second, the value of $\lambda_D=d D(n)/dn$ can only be extracted from the non-linear fit of the data }  (see Supplemental Material).
%

%

\textit{Bjorken expansion: } 
A simple dynamical system to study the effect of a UV-regulator on diffusive fluctuations is the Bjorken flow, which is a hydrodynamic model for the longitudinal expansion of the QGP.

The effect of non-linear fluctuations in Bjorken flow has been studied in Ref.~\cite{Akamatsu:2016llw}. Developing a set of hydro-kinetic equations, the first fractional power correction to the longitudinal pressure at late times  was found as $\propto 1/(\tau_{p} T)^{3/2}$($\tau_{p}$ is the proper time and $1/\tau_{p}$ the expansion rate of the Bjorken flow). If the expanding flow has a $U(1)$ charge, the fluctuations also result in a fractional power correction of the $U(1)$ current. This is the consequence of non-linear mode-coupling between the currents and the hydrodynamic variables \cite{Martinez:2018wia}. In contrast to previous studies, our model also features a UV-regulator, {i.e.}~$\tau$.
Without deriving hydro-kinetic equations, we estimate the effect of the UV-regulator on the late-time non-linear correction to the single charge density,  $\Delta\langle n(\tau_p) \rangle$, in ``non-fluctuating" Bjorken flow. 
In contrast to \cite{Martinez:2018wia}, the effect 
comes from the self-interactions of $n$. 
We find (Supplemental Material)
\begin{equation}\label{Bjorken}
\Delta\langle n(\tau_{p})\rangle =
 a\,T\chi^2\mu\,\frac{\lambda_D^2}{D^2}\frac{1}{(D \tau_p)^{3/2}}\left(1- \frac{11}{8}\frac{\tau}{\tau_p}+\cdots\right) \, .
	\end{equation} 
The leading correction is similar to Eq.~(78a) in~\cite{Martinez:2018wia}. In fact this term can be found through the work of~\cite{Chen-Lin:2018kfl,Chao:2020kcf}. However, the sub-leading term is found entirely from the theory proposed in this letter, representing the effect of the UV-regulator. 
It is important but if $\tau\ll \tau_{p}$ smaller than the leading non-linear effects. In order to make the above results more accurate in a more realistic scenario, it would be interesting to investigate the effect of the UV-regulator on the hydro-kinetic setup of~\cite{Martinez:2018wia}. 

\nw{Another aspect of our theory of non-linear diffusion, worthy of future exploration, is its application near a critical point, such as for example the critical point expected in the QCD phase diagram.  
Near such a critical point charge fluctuations, the correlation length, and the relaxation time of our UV-mode all become large~\cite{hohenberg1977theory}, such that it has to be included into the long-wavelength description~\cite{Stephanov:2017ghc}, making our non-linear theory of diffusion relevant to the search for the critical point~\cite{Stephanov:1998dy,Du:2020bxp,Stephanov:2008qz}.} 

\so{
Fluctuations in the UV-regulated theory of diffusion are also important near a critical point. Let us suppose a simple diffusive charge near the critical point. In the dynamic model H~\cite{hohenberg1977theory}, }
\begin{equation*}
	\so{D\sim\xi^{-1}\,,\,\,\,\,\,\,\,\,\chi\sim \xi^2\,,\,\,\,\,\,\,\,\,\,\kappa\sim\xi \, ,}
\end{equation*}
\so{with $\xi$ being the correlation length. It turns out that $\tau\sim\xi^2$~\cite{Du:2021zqz}.
Near the critical point $\xi$ becomes large, so does the relaxation time $\tau$. As a result, the UV mode of our theory becomes a slow mode and then has to be included even in the long-wavelength limit~\cite{Stephanov:2017ghc}. 
On the other hand, we find that $\Sigma_3\sim\xi^{9/2}$, becoming large near the critical point. This shows the need to consider non-linearities discussed in this paper. \\
Within a QGP droplet passing by the conjectured QCD critical point, diffusion is coupled to bulk dynamics \cite{Du:2020bxp}. The 
model discussed in this paper then needs to be coupled with relativistic hydrodynamic fluctuations. 
The effect of interactions and self-interactions in non-linear MIS theory on the final proton multiplicity cumulants should then be investigated~\cite{Stephanov:2008qz}, and is relevant to the search for the critical point~\cite{Stephanov:1998dy}. 
}

\begin{acknowledgments}

	We thank P.~Kovtun and A.~Davody for helpful discussions. For comments on the draft we thank L.~Delacr{\'e}taz, P.~Glorioso, and E.~van Heumen. 
{M.K.} was supported, by the {U.S.}~Department of Energy grant DE-SC0012447. {M.K.} thanks the Universit{\"a}t W{\"u}rzburg and the APCTP where part of this work was carried out. {N.A.} was supported by grant number 561119208 “Double First Class” start-up funding of Lanzhou University, China.
\end{acknowledgments}


\begin{thebibliography}{33}%
\makeatletter
\providecommand \@ifxundefined [1]{%
 \@ifx{#1\undefined}
}%
\providecommand \@ifnum [1]{%
 \ifnum #1\expandafter \@firstoftwo
 \else \expandafter \@secondoftwo
 \fi
}%
\providecommand \@ifx [1]{%
 \ifx #1\expandafter \@firstoftwo
 \else \expandafter \@secondoftwo
 \fi
}%
\providecommand \natexlab [1]{#1}%
\providecommand \enquote  [1]{``#1''}%
\providecommand \bibnamefont  [1]{#1}%
\providecommand \bibfnamefont [1]{#1}%
\providecommand \citenamefont [1]{#1}%
\providecommand \href@noop [0]{\@secondoftwo}%
\providecommand \href [0]{\begingroup \@sanitize@url \@href}%
\providecommand \@href[1]{\@@startlink{#1}\@@href}%
\providecommand \@@href[1]{\endgroup#1\@@endlink}%
\providecommand \@sanitize@url [0]{\catcode `\\12\catcode `\$12\catcode
  `\&12\catcode `\#12\catcode `\^12\catcode `\_12\catcode `\%12\relax}%
\providecommand \@@startlink[1]{}%
\providecommand \@@endlink[0]{}%
\providecommand \url  [0]{\begingroup\@sanitize@url \@url }%
\providecommand \@url [1]{\endgroup\@href {#1}{\urlprefix }}%
\providecommand \urlprefix  [0]{URL }%
\providecommand \Eprint [0]{\href }%
\providecommand \doibase [0]{http://dx.doi.org/}%
\providecommand \selectlanguage [0]{\@gobble}%
\providecommand \bibinfo  [0]{\@secondoftwo}%
\providecommand \bibfield  [0]{\@secondoftwo}%
\providecommand \translation [1]{[#1]}%
\providecommand \BibitemOpen [0]{}%
\providecommand \bibitemStop [0]{}%
\providecommand \bibitemNoStop [0]{.\EOS\space}%
\providecommand \EOS [0]{\spacefactor3000\relax}%
\providecommand \BibitemShut  [1]{\csname bibitem#1\endcsname}%
\let\auto@bib@innerbib\@empty
\bibitem [{\citenamefont {Crossley}\ \emph {et~al.}(2017)\citenamefont
  {Crossley}, \citenamefont {Glorioso},\ and\ \citenamefont
  {Liu}}]{Crossley:2015evo}%
  \BibitemOpen
  \bibfield  {author} {\bibinfo {author} {\bibfnamefont {M.}~\bibnamefont
  {Crossley}}, \bibinfo {author} {\bibfnamefont {P.}~\bibnamefont {Glorioso}},
  \ and\ \bibinfo {author} {\bibfnamefont {H.}~\bibnamefont {Liu}},\ }\href
  {\doibase 10.1007/JHEP09(2017)095} {\bibfield  {journal} {\bibinfo  {journal}
  {JHEP}\ }\textbf {\bibinfo {volume} {09}},\ \bibinfo {pages} {095} (\bibinfo
  {year} {2017})},\ \Eprint {http://arxiv.org/abs/1511.03646} {arXiv:1511.03646
  [hep-th]} \BibitemShut {NoStop}%
\bibitem [{\citenamefont {Jensen}\ \emph {et~al.}(2018)\citenamefont {Jensen},
  \citenamefont {Marjieh}, \citenamefont {Pinzani-Fokeeva},\ and\ \citenamefont
  {Yarom}}]{Jensen:2018hse}%
  \BibitemOpen
  \bibfield  {author} {\bibinfo {author} {\bibfnamefont {K.}~\bibnamefont
  {Jensen}}, \bibinfo {author} {\bibfnamefont {R.}~\bibnamefont {Marjieh}},
  \bibinfo {author} {\bibfnamefont {N.}~\bibnamefont {Pinzani-Fokeeva}}, \ and\
  \bibinfo {author} {\bibfnamefont {A.}~\bibnamefont {Yarom}},\ }\href@noop {}
  {\  (\bibinfo {year} {2018})},\ \Eprint {http://arxiv.org/abs/1804.04654}
  {arXiv:1804.04654 [hep-th]} \BibitemShut {NoStop}%
\bibitem [{\citenamefont {Haehl}\ \emph {et~al.}(2015)\citenamefont {Haehl},
  \citenamefont {Loganayagam},\ and\ \citenamefont
  {Rangamani}}]{Haehl:2015pja}%
  \BibitemOpen
  \bibfield  {author} {\bibinfo {author} {\bibfnamefont {F.~M.}\ \bibnamefont
  {Haehl}}, \bibinfo {author} {\bibfnamefont {R.}~\bibnamefont {Loganayagam}},
  \ and\ \bibinfo {author} {\bibfnamefont {M.}~\bibnamefont {Rangamani}},\
  }\href {\doibase 10.1007/JHEP05(2015)060} {\bibfield  {journal} {\bibinfo
  {journal} {JHEP}\ }\textbf {\bibinfo {volume} {05}},\ \bibinfo {pages} {060}
  (\bibinfo {year} {2015})},\ \Eprint {http://arxiv.org/abs/1502.00636}
  {arXiv:1502.00636 [hep-th]} \BibitemShut {NoStop}%
\bibitem [{\citenamefont {Liu}\ and\ \citenamefont
  {Glorioso}(2018)}]{Glorioso:2018wxw}%
  \BibitemOpen
  \bibfield  {author} {\bibinfo {author} {\bibfnamefont {H.}~\bibnamefont
  {Liu}}\ and\ \bibinfo {author} {\bibfnamefont {P.}~\bibnamefont {Glorioso}},\
  }\bibfield  {booktitle} { {\bibinfo {booktitle} {{Proceedings,
  Theoretical Advanced Study Institute in Elementary Particle Physics: Physics
  at the Fundamental Frontier (TASI 2017): Boulder, CO, USA, June 5-30,
  2017}}},\ }\href {\doibase 10.22323/1.305.0008} {\bibfield  {journal}
  {\bibinfo  {journal} {PoS}\ }\textbf {\bibinfo {volume} {TASI2017}},\
  \bibinfo {pages} {008} (\bibinfo {year} {2018})},\ \Eprint
  {http://arxiv.org/abs/1805.09331} {arXiv:1805.09331 [hep-th]} \BibitemShut
  {NoStop}%
\bibitem [{Note1()}]{Note1}%
  \BibitemOpen
  \bibinfo {note} {This was discussed in \cite
  {Kovtun:2014nsa,Chen-Lin:2018kfl,Chao:2020kcf}. While \cite
  {Chen-Lin:2018kfl} focuses on density fluctuations in the framework of
  Schwinger-Keldysh, \cite {Chao:2020kcf} studies multiplicative noise, arisen
  from such non-linear dependence in the context of stochastic fluid dynamics.
  See also~\cite {Chao:2023kvz}.}\BibitemShut {Stop}%
\bibitem [{\citenamefont {Chen-Lin}\ \emph {et~al.}(2019)\citenamefont
  {Chen-Lin}, \citenamefont {Delacr\'etaz},\ and\ \citenamefont
  {Hartnoll}}]{Chen-Lin:2018kfl}%
  \BibitemOpen
  \bibfield  {author} {\bibinfo {author} {\bibfnamefont {X.}~\bibnamefont
  {Chen-Lin}}, \bibinfo {author} {\bibfnamefont {L.~V.}\ \bibnamefont
  {Delacr\'etaz}}, \ and\ \bibinfo {author} {\bibfnamefont {S.~A.}\
  \bibnamefont {Hartnoll}},\ }\href {\doibase 10.1103/PhysRevLett.122.091602}
  {\bibfield  {journal} {\bibinfo  {journal} {Phys. Rev. Lett.}\ }\textbf
  {\bibinfo {volume} {122}},\ \bibinfo {pages} {091602} (\bibinfo {year}
  {2019})},\ \Eprint {http://arxiv.org/abs/1811.12540} {arXiv:1811.12540
  [hep-th]} \BibitemShut {NoStop}%
\bibitem [{\citenamefont {Chao}\ and\ \citenamefont
  {Schaefer}(2021)}]{Chao:2020kcf}%
  \BibitemOpen
  \bibfield  {author} {\bibinfo {author} {\bibfnamefont {J.}~\bibnamefont
  {Chao}}\ and\ \bibinfo {author} {\bibfnamefont {T.}~\bibnamefont
  {Schaefer}},\ }\href {\doibase 10.1007/JHEP01(2021)071} {\bibfield  {journal}
  {\bibinfo  {journal} {JHEP}\ }\textbf {\bibinfo {volume} {01}},\ \bibinfo
  {pages} {071} (\bibinfo {year} {2021})},\ \Eprint
  {http://arxiv.org/abs/2008.01269} {arXiv:2008.01269 [hep-th]} \BibitemShut
  {NoStop}%
\bibitem [{\citenamefont {Kovtun}\ and\ \citenamefont
  {Yaffe}(2003)}]{Kovtun:2003vj}%
  \BibitemOpen
  \bibfield  {author} {\bibinfo {author} {\bibfnamefont {P.}~\bibnamefont
  {Kovtun}}\ and\ \bibinfo {author} {\bibfnamefont {L.~G.}\ \bibnamefont
  {Yaffe}},\ }\href {\doibase 10.1103/PhysRevD.68.025007} {\bibfield  {journal}
  {\bibinfo  {journal} {Phys. Rev. D}\ }\textbf {\bibinfo {volume} {68}},\
  \bibinfo {pages} {025007} (\bibinfo {year} {2003})},\ \Eprint
  {http://arxiv.org/abs/hep-th/0303010} {arXiv:hep-th/0303010} \BibitemShut
  {NoStop}%
\bibitem [{\citenamefont {Kovtun}\ \emph {et~al.}(2011)\citenamefont {Kovtun},
  \citenamefont {Moore},\ and\ \citenamefont {Romatschke}}]{Kovtun:2011np}%
  \BibitemOpen
  \bibfield  {author} {\bibinfo {author} {\bibfnamefont {P.}~\bibnamefont
  {Kovtun}}, \bibinfo {author} {\bibfnamefont {G.~D.}\ \bibnamefont {Moore}}, \
  and\ \bibinfo {author} {\bibfnamefont {P.}~\bibnamefont {Romatschke}},\
  }\href {\doibase 10.1103/PhysRevD.84.025006} {\bibfield  {journal} {\bibinfo
  {journal} {Phys. Rev. D}\ }\textbf {\bibinfo {volume} {84}},\ \bibinfo
  {pages} {025006} (\bibinfo {year} {2011})},\ \Eprint
  {http://arxiv.org/abs/1104.1586} {arXiv:1104.1586 [hep-ph]} \BibitemShut
  {NoStop}%
\bibitem [{\citenamefont {Kadanoff}\ and\ \citenamefont
  {Martin}(1963)}]{kadanoff1963hydrodynamic}%
  \BibitemOpen
  \bibfield  {author} {\bibinfo {author} {\bibfnamefont {L.~P.}\ \bibnamefont
  {Kadanoff}}\ and\ \bibinfo {author} {\bibfnamefont {P.~C.}\ \bibnamefont
  {Martin}},\ }\href@noop {} {\bibfield  {journal} {\bibinfo  {journal} {Annals
  of Physics}\ }\textbf {\bibinfo {volume} {24}},\ \bibinfo {pages} {419}
  (\bibinfo {year} {1963})}\BibitemShut {NoStop}%
\bibitem [{\citenamefont {Brown~et al}(2019)}]{Brown:2018}%
  \BibitemOpen
  \bibfield  {author} {\bibinfo {author} {\bibfnamefont {P.~T.}\ \bibnamefont
  {Brown~et al}},\ }\href {\doibase https://doi.org/10.1126/science.aat4134}
  {\bibfield  {journal} {\bibinfo  {journal} {Science}\ }\textbf {\bibinfo
  {volume} {363}},\ \bibinfo {pages} {379} (\bibinfo {year} {2019})},\ \Eprint
  {http://arxiv.org/abs/1802.09456} {arXiv:1802.09456 [cond-mat.quant-gas]}
  \BibitemShut {NoStop}%
\bibitem [{\citenamefont {Hartnoll}\ and\ \citenamefont
  {Mackenzie}(2022)}]{Hartnoll:2021ydi}%
  \BibitemOpen
  \bibfield  {author} {\bibinfo {author} {\bibfnamefont {S.~A.}\ \bibnamefont
  {Hartnoll}}\ and\ \bibinfo {author} {\bibfnamefont {A.~P.}\ \bibnamefont
  {Mackenzie}},\ }\href {\doibase 10.1103/RevModPhys.94.041002} {\bibfield
  {journal} {\bibinfo  {journal} {Rev. Mod. Phys.}\ }\textbf {\bibinfo {volume}
  {94}},\ \bibinfo {pages} {041002} (\bibinfo {year} {2022})},\ \Eprint
  {http://arxiv.org/abs/2107.07802} {arXiv:2107.07802 [cond-mat.str-el]}
  \BibitemShut {NoStop}%
\bibitem [{Note2()}]{Note2}%
  \BibitemOpen
  \bibinfo {note} {When the system is not static, flow dependent timescales are
  also involved in thermalization. Examples are the weakly coupled system
  of~\cite {Behtash:2019txb} under the Bjorken flow, and~\cite {Bazow:2015dha}
  under the FLRW expansion.}\BibitemShut {Stop}%
\bibitem [{\citenamefont {Ernst}\ \emph {et~al.}(1970)\citenamefont {Ernst},
  \citenamefont {Hauge},\ and\ \citenamefont {van
  Leeuwen}}]{ernst1970asymptotic}%
  \BibitemOpen
  \bibfield  {author} {\bibinfo {author} {\bibfnamefont {M.~H.}\ \bibnamefont
  {Ernst}}, \bibinfo {author} {\bibfnamefont {E.~H.}\ \bibnamefont {Hauge}}, \
  and\ \bibinfo {author} {\bibfnamefont {J.~M.~J.}\ \bibnamefont {van
  Leeuwen}},\ }\href@noop {} {\bibfield  {journal} {\bibinfo  {journal}
  {Physical Review Letters}\ }\textbf {\bibinfo {volume} {25}},\ \bibinfo
  {pages} {1254} (\bibinfo {year} {1970})}\BibitemShut {NoStop}%
\bibitem [{\citenamefont {Alder}\ and\ \citenamefont
  {Wainwright}(1970)}]{alder1970decay}%
  \BibitemOpen
  \bibfield  {author} {\bibinfo {author} {\bibfnamefont {B.~J.}\ \bibnamefont
  {Alder}}\ and\ \bibinfo {author} {\bibfnamefont {T.~E.}\ \bibnamefont
  {Wainwright}},\ }\href@noop {} {\bibfield  {journal} {\bibinfo  {journal}
  {Physical Review A}\ }\textbf {\bibinfo {volume} {1}},\ \bibinfo {pages} {18}
  (\bibinfo {year} {1970})}\BibitemShut {NoStop}%
\bibitem [{Note3()}]{Note3}%
  \BibitemOpen
  \bibinfo {note} {We focus on static systems. Systems in a dynamical
  background may not exhibit such exponential behavior; for example the Gubser
  flow~\cite {Gubser:2010ze}.}\BibitemShut {Stop}%
\bibitem [{\citenamefont {Martin}\ \emph {et~al.}(1973)\citenamefont {Martin},
  \citenamefont {Siggia},\ and\ \citenamefont {Rose}}]{martin1973statistical}%
  \BibitemOpen
  \bibfield  {author} {\bibinfo {author} {\bibfnamefont {P.~C.}\ \bibnamefont
  {Martin}}, \bibinfo {author} {\bibfnamefont {E.~D.}\ \bibnamefont {Siggia}},
  \ and\ \bibinfo {author} {\bibfnamefont {H.~A.}\ \bibnamefont {Rose}},\
  }\href@noop {} {\bibfield  {journal} {\bibinfo  {journal} {Physical Review
  A}\ }\textbf {\bibinfo {volume} {8}},\ \bibinfo {pages} {423} (\bibinfo
  {year} {1973})}\BibitemShut {NoStop}%
\bibitem [{\citenamefont {Kovtun}(2015)}]{Kovtun:2014nsa}%
  \BibitemOpen
  \bibfield  {author} {\bibinfo {author} {\bibfnamefont {P.}~\bibnamefont
  {Kovtun}},\ }\href {\doibase 10.1088/1751-8113/48/26/265002} {\bibfield
  {journal} {\bibinfo  {journal} {J. Phys. A}\ }\textbf {\bibinfo {volume}
  {48}},\ \bibinfo {pages} {265002} (\bibinfo {year} {2015})},\ \Eprint
  {http://arxiv.org/abs/1407.0690} {arXiv:1407.0690 [cond-mat.stat-mech]}
  \BibitemShut {NoStop}%
\bibitem [{\citenamefont {Michailidis}\ \emph {et~al.}(2023)\citenamefont
  {Michailidis}, \citenamefont {Abanin},\ and\ \citenamefont
  {Delacr\'etaz}}]{Michailidis:2023mkd}%
  \BibitemOpen
  \bibfield  {author} {\bibinfo {author} {\bibfnamefont {A.~A.}\ \bibnamefont
  {Michailidis}}, \bibinfo {author} {\bibfnamefont {D.~A.}\ \bibnamefont
  {Abanin}}, \ and\ \bibinfo {author} {\bibfnamefont {L.~V.}\ \bibnamefont
  {Delacr\'etaz}},\ }\href@noop {} {\  (\bibinfo {year} {2023})},\ \Eprint
  {http://arxiv.org/abs/2310.10564} {arXiv:2310.10564 [cond-mat.stat-mech]}
  \BibitemShut {NoStop}%
\bibitem [{\citenamefont {Heller}\ and\ \citenamefont
  {Spalinski}(2015)}]{Heller:2015dha}%
  \BibitemOpen
  \bibfield  {author} {\bibinfo {author} {\bibfnamefont {M.~P.}\ \bibnamefont
  {Heller}}\ and\ \bibinfo {author} {\bibfnamefont {M.}~\bibnamefont
  {Spalinski}},\ }\href {\doibase 10.1103/PhysRevLett.115.072501} {\bibfield
  {journal} {\bibinfo  {journal} {Phys. Rev. Lett.}\ }\textbf {\bibinfo
  {volume} {115}},\ \bibinfo {pages} {072501} (\bibinfo {year} {2015})},\
  \Eprint {http://arxiv.org/abs/1503.07514} {arXiv:1503.07514 [hep-th]}
  \BibitemShut {NoStop}%
\bibitem [{\citenamefont {Delacretaz}(2020)}]{Delacretaz:2020nit}%
  \BibitemOpen
  \bibfield  {author} {\bibinfo {author} {\bibfnamefont {L.~V.}\ \bibnamefont
  {Delacretaz}},\ }\href {\doibase 10.21468/SciPostPhys.9.3.034} {\bibfield
  {journal} {\bibinfo  {journal} {SciPost Phys.}\ }\textbf {\bibinfo {volume}
  {9}},\ \bibinfo {pages} {034} (\bibinfo {year} {2020})},\ \Eprint
  {http://arxiv.org/abs/2006.01139} {arXiv:2006.01139 [hep-th]} \BibitemShut
  {NoStop}%
\bibitem [{\citenamefont {Hohenberg}\ and\ \citenamefont
  {Halperin}(1977)}]{hohenberg1977theory}%
  \BibitemOpen
  \bibfield  {author} {\bibinfo {author} {\bibfnamefont {P.~C.}\ \bibnamefont
  {Hohenberg}}\ and\ \bibinfo {author} {\bibfnamefont {B.~I.}\ \bibnamefont
  {Halperin}},\ }\href {\doibase 10.1103/RevModPhys.49.435} {\bibfield
  {journal} {\bibinfo  {journal} {Rev. Mod. Phys.}\ }\textbf {\bibinfo {volume}
  {49}},\ \bibinfo {pages} {435} (\bibinfo {year} {1977})}\BibitemShut
  {NoStop}%
\bibitem [{\citenamefont {Stephanov}\ and\ \citenamefont
  {Yin}(2018)}]{Stephanov:2017ghc}%
  \BibitemOpen
  \bibfield  {author} {\bibinfo {author} {\bibfnamefont {M.}~\bibnamefont
  {Stephanov}}\ and\ \bibinfo {author} {\bibfnamefont {Y.}~\bibnamefont
  {Yin}},\ }\href {\doibase 10.1103/PhysRevD.98.036006} {\bibfield  {journal}
  {\bibinfo  {journal} {Phys. Rev. D}\ }\textbf {\bibinfo {volume} {98}},\
  \bibinfo {pages} {036006} (\bibinfo {year} {2018})},\ \Eprint
  {http://arxiv.org/abs/1712.10305} {arXiv:1712.10305 [nucl-th]} \BibitemShut
  {NoStop}%
\bibitem [{\citenamefont {Du}\ \emph {et~al.}(2020)\citenamefont {Du},
  \citenamefont {Heinz}, \citenamefont {Rajagopal},\ and\ \citenamefont
  {Yin}}]{Du:2020bxp}%
  \BibitemOpen
  \bibfield  {author} {\bibinfo {author} {\bibfnamefont {L.}~\bibnamefont
  {Du}}, \bibinfo {author} {\bibfnamefont {U.}~\bibnamefont {Heinz}}, \bibinfo
  {author} {\bibfnamefont {K.}~\bibnamefont {Rajagopal}}, \ and\ \bibinfo
  {author} {\bibfnamefont {Y.}~\bibnamefont {Yin}},\ }\href {\doibase
  10.1103/PhysRevC.102.054911} {\bibfield  {journal} {\bibinfo  {journal}
  {Phys. Rev. C}\ }\textbf {\bibinfo {volume} {102}},\ \bibinfo {pages}
  {054911} (\bibinfo {year} {2020})},\ \Eprint
  {http://arxiv.org/abs/2004.02719} {arXiv:2004.02719 [nucl-th]} \BibitemShut
  {NoStop}%
\bibitem [{\citenamefont {Stephanov}(2009)}]{Stephanov:2008qz}%
  \BibitemOpen
  \bibfield  {author} {\bibinfo {author} {\bibfnamefont {M.~A.}\ \bibnamefont
  {Stephanov}},\ }\href {\doibase 10.1103/PhysRevLett.102.032301} {\bibfield
  {journal} {\bibinfo  {journal} {Phys. Rev. Lett.}\ }\textbf {\bibinfo
  {volume} {102}},\ \bibinfo {pages} {032301} (\bibinfo {year} {2009})},\
  \Eprint {http://arxiv.org/abs/0809.3450} {arXiv:0809.3450 [hep-ph]}
  \BibitemShut {NoStop}%
\bibitem [{\citenamefont {Stephanov}\ \emph {et~al.}(1998)\citenamefont
  {Stephanov}, \citenamefont {Rajagopal},\ and\ \citenamefont
  {Shuryak}}]{Stephanov:1998dy}%
  \BibitemOpen
  \bibfield  {author} {\bibinfo {author} {\bibfnamefont {M.~A.}\ \bibnamefont
  {Stephanov}}, \bibinfo {author} {\bibfnamefont {K.}~\bibnamefont
  {Rajagopal}}, \ and\ \bibinfo {author} {\bibfnamefont {E.~V.}\ \bibnamefont
  {Shuryak}},\ }\href {\doibase 10.1103/PhysRevLett.81.4816} {\bibfield
  {journal} {\bibinfo  {journal} {Phys. Rev. Lett.}\ }\textbf {\bibinfo
  {volume} {81}},\ \bibinfo {pages} {4816} (\bibinfo {year} {1998})},\ \Eprint
  {http://arxiv.org/abs/hep-ph/9806219} {arXiv:hep-ph/9806219} \BibitemShut
  {NoStop}%
\bibitem [{\citenamefont {Akamatsu}\ \emph {et~al.}(2017)\citenamefont
  {Akamatsu}, \citenamefont {Mazeliauskas},\ and\ \citenamefont
  {Teaney}}]{Akamatsu:2016llw}%
  \BibitemOpen
  \bibfield  {author} {\bibinfo {author} {\bibfnamefont {Y.}~\bibnamefont
  {Akamatsu}}, \bibinfo {author} {\bibfnamefont {A.}~\bibnamefont
  {Mazeliauskas}}, \ and\ \bibinfo {author} {\bibfnamefont {D.}~\bibnamefont
  {Teaney}},\ }\href {\doibase 10.1103/PhysRevC.95.014909} {\bibfield
  {journal} {\bibinfo  {journal} {Phys. Rev. C}\ }\textbf {\bibinfo {volume}
  {95}},\ \bibinfo {pages} {014909} (\bibinfo {year} {2017})},\ \Eprint
  {http://arxiv.org/abs/1606.07742} {arXiv:1606.07742 [nucl-th]} \BibitemShut
  {NoStop}%
\bibitem [{\citenamefont {Martinez}\ and\ \citenamefont
  {Sch\"afer}(2019)}]{Martinez:2018wia}%
  \BibitemOpen
  \bibfield  {author} {\bibinfo {author} {\bibfnamefont {M.}~\bibnamefont
  {Martinez}}\ and\ \bibinfo {author} {\bibfnamefont {T.}~\bibnamefont
  {Sch\"afer}},\ }\href {\doibase 10.1103/PhysRevC.99.054902} {\bibfield
  {journal} {\bibinfo  {journal} {Phys. Rev. C}\ }\textbf {\bibinfo {volume}
  {99}},\ \bibinfo {pages} {054902} (\bibinfo {year} {2019})},\ \Eprint
  {http://arxiv.org/abs/1812.05279} {arXiv:1812.05279 [hep-th]} \BibitemShut
  {NoStop}%
\bibitem [{\citenamefont {Chao}\ and\ \citenamefont
  {Schaefer}(2023)}]{Chao:2023kvz}%
  \BibitemOpen
  \bibfield  {author} {\bibinfo {author} {\bibfnamefont {J.}~\bibnamefont
  {Chao}}\ and\ \bibinfo {author} {\bibfnamefont {T.}~\bibnamefont
  {Schaefer}},\ }\href {\doibase 10.1007/JHEP06(2023)057} {\bibfield  {journal}
  {\bibinfo  {journal} {JHEP}\ }\textbf {\bibinfo {volume} {06}},\ \bibinfo
  {pages} {057} (\bibinfo {year} {2023})},\ \Eprint
  {http://arxiv.org/abs/2302.00720} {arXiv:2302.00720 [hep-ph]} \BibitemShut
  {NoStop}%
\bibitem [{\citenamefont {Behtash}\ \emph {et~al.}(2019)\citenamefont
  {Behtash}, \citenamefont {Kamata}, \citenamefont {Martinez},\ and\
  \citenamefont {Shi}}]{Behtash:2019txb}%
  \BibitemOpen
  \bibfield  {author} {\bibinfo {author} {\bibfnamefont {A.}~\bibnamefont
  {Behtash}}, \bibinfo {author} {\bibfnamefont {S.}~\bibnamefont {Kamata}},
  \bibinfo {author} {\bibfnamefont {M.}~\bibnamefont {Martinez}}, \ and\
  \bibinfo {author} {\bibfnamefont {H.}~\bibnamefont {Shi}},\ }\href {\doibase
  10.1103/PhysRevD.99.116012} {\bibfield  {journal} {\bibinfo  {journal} {Phys.
  Rev. D}\ }\textbf {\bibinfo {volume} {99}},\ \bibinfo {pages} {116012}
  (\bibinfo {year} {2019})},\ \Eprint {http://arxiv.org/abs/1901.08632}
  {arXiv:1901.08632 [hep-th]} \BibitemShut {NoStop}%
\bibitem [{\citenamefont {Bazow}\ \emph {et~al.}(2016)\citenamefont {Bazow},
  \citenamefont {Denicol}, \citenamefont {Heinz}, \citenamefont {Martinez},\
  and\ \citenamefont {Noronha}}]{Bazow:2015dha}%
  \BibitemOpen
  \bibfield  {author} {\bibinfo {author} {\bibfnamefont {D.}~\bibnamefont
  {Bazow}}, \bibinfo {author} {\bibfnamefont {G.~S.}\ \bibnamefont {Denicol}},
  \bibinfo {author} {\bibfnamefont {U.}~\bibnamefont {Heinz}}, \bibinfo
  {author} {\bibfnamefont {M.}~\bibnamefont {Martinez}}, \ and\ \bibinfo
  {author} {\bibfnamefont {J.}~\bibnamefont {Noronha}},\ }\href {\doibase
  10.1103/PhysRevLett.116.022301} {\bibfield  {journal} {\bibinfo  {journal}
  {Phys. Rev. Lett.}\ }\textbf {\bibinfo {volume} {116}},\ \bibinfo {pages}
  {022301} (\bibinfo {year} {2016})},\ \Eprint
  {http://arxiv.org/abs/1507.07834} {arXiv:1507.07834 [hep-ph]} \BibitemShut
  {NoStop}%
\bibitem [{\citenamefont {Gubser}(2010)}]{Gubser:2010ze}%
  \BibitemOpen
  \bibfield  {author} {\bibinfo {author} {\bibfnamefont {S.~S.}\ \bibnamefont
  {Gubser}},\ }\href {\doibase 10.1103/PhysRevD.82.085027} {\bibfield
  {journal} {\bibinfo  {journal} {Phys. Rev. D}\ }\textbf {\bibinfo {volume}
  {82}},\ \bibinfo {pages} {085027} (\bibinfo {year} {2010})},\ \Eprint
  {http://arxiv.org/abs/1006.0006} {arXiv:1006.0006 [hep-th]} \BibitemShut
  {NoStop}%
\end{thebibliography}%

\end{document}


\title{Supplemental Material}

%
%


\maketitle

\onecolumngrid

\setcounter{equation}{16}
\setcounter{figure}{3}

\section{Correlation function in classical and quantum limits}
\label{correlation_functions}
From the linear equation ~(2),
one can easily compute the correlation functions between the fluctuations of the density.
Following \cite{landau2013statistical}, the equations of one-point functions given in~(1) 
can be promoted to equations for the \textit{correlation functions}:
\begin{equation}\label{coupled_correlations}
	\begin{split}
		\partial_t \langle  n(t,\textbf{x})  n(0,\textbf{0})\rangle+\boldsymbol{\nabla}\cdot \langle \textbf{J}(t,\textbf{x})  n(0,\textbf{0})\rangle=&\,0\\
		\tau\, \partial_t \langle \textbf{J}(t,\textbf{x})  n(0,\textbf{0})\rangle+\langle \textbf{J}(t,\textbf{x})  n(0,\textbf{0})\rangle+ D \,\boldsymbol{\nabla}\langle  n(t,\textbf{x})  n(0,\textbf{0})\rangle=&\,0
	\end{split}
\end{equation}
\textit{Note that the above averages are over background thermodynamic fluctuations, the so-called noise fields.} 
In the absence of correlation at non-zero distances, the equal-time correlation functions reduce to those being familiar from thermodynamics.
The initial boundary condition  is then the vanishing of the equal-time correlation function of the $n$ and $\textbf{J}$ together with 
\begin{equation}\label{}
	\langle  n(\textbf{x}_1) n(\textbf{x}_2)\rangle=\,T\,\chi\,\delta (\textbf{x}_1-\textbf{x}_2)\,.
\end{equation}
Let us define 
$\langle  n_{\omega \textbf{k}} n_{\omega' \textbf{k}'}\rangle=\,(2\pi)^4\delta(\omega+\omega')\delta(\textbf{k}+\textbf{k}')\langle nn\rangle_{\omega\textbf{k}}$. 
Then by a one-sided Fourier transformation with respect to time and complete transformation with respect to spatial coordinates, 
equations \eqref{coupled_correlations} simplify to
\begin{equation}
	\begin{split}
		- T\chi -i \omega  \, \langle nn\rangle_{\omega\textbf{k}}^{(+)}+i \textbf{k}\cdot \langle\textbf{J}\, n\rangle_{\omega\textbf{k}}^{(+)}=&\,0\\
		-i \omega \tau\langle\textbf{J}\, n\rangle_{\omega\textbf{k}}^{(+)}+\langle\textbf{J}\, n\rangle_{\omega\textbf{k}}^{(+)}+i D \textbf{k}\cdot \langle nn\rangle_{\omega\textbf{k}}^{(+)} =&\,0
	\end{split}
\end{equation}
where $\langle nn\rangle_{\omega\textbf{k}}^{(+)}=\,\int_0^{+\infty}dt\int\langle  n(t,\textbf{x})  n(0,\textbf{0})\rangle e^{-i (\textbf{k}\cdot\textbf{x}-\omega t)}\,d^3\text{x}$. 
The correlation function then reads 	
\begin{equation}\label{correlation}
G_{nn}(\omega, \textbf{k})\equiv\langle nn\rangle_{\omega\textbf{k}}=\,2\left(\frac{\omega}{2T}\right)\coth\left(\frac{\omega}{2T}\right)\ \text{Re}\, \langle nn\rangle_{\omega\textbf{k}}^{(+)}=\,\frac{\left(\frac{\omega}{2T}\right)\coth\left(\frac{\omega}{2T}\right)2\,T\chi\,D\textbf{k}^2}{\omega^2+(\tau \omega^2 -D \textbf{k}^2)^2}
\end{equation}
Note that the above first equality is the quantum mechanical version of the relation between the correlation function and its one-sided Fourier transform~\cite{landau2013statistical}. Since we want to add a UV regulator to the diffusion theory, $\omega$ can be of order of $T$ and consequently the presence of such quantum effects are inevitable. In the classical limit $\omega\ll T$, and then the expression  $\left(\frac{\omega}{2T}\right)\coth\left(\frac{\omega}{2T}\right)$ reduces to unity. Deviations from the classical limit may be written as
\begin{equation}\label{QM_to_Classic}
	Q(\omega)\equiv\left(\frac{\omega}{2T}\right)\coth\left(\frac{\omega}{2T}\right)=\,1+\frac{1}{12}\left(\frac{\omega}{T}\right)^2-\frac{1}{720}\left(\frac{\omega}{T}\right)^4+ \mathcal{O}\left(\frac{\omega}{T}\right)^6\,.
\end{equation}
\textit{Is the above Taylor expansion convergent?} The answer is not, because the left hand side has a set of branch point singularities in the complex frequency plane at $\omega=\pm i 2 \pi n T,\,\,n\ne0$. The domain of convergence of the above expansion is then limited to $|\omega|<2\pi T$. Then the next question is \textit{how much large the momentum cutoff $\Lambda_{\text{EFT}}$ can be considered in our setup?} Based on the above discussion, in order to continue to use the Taylor expansion \eqref{QM_to_Classic}, we may take $\Lambda_{\text{EFT}}\lesssim 2\pi T$. However, the larger the value of $\Lambda_{\text{EFT}}$, the slower the series in \eqref{QM_to_Classic} converges. For this reason, we take $\Lambda_{\text{EFT}}\simeq3T$. In this way, the first three terms in the series are sufficient to obtain an accurate result in perturbation theory with less than $2\%$ error. 
On the other hand, the gap mode $\omega_2$ needs to belong to the spectrum; $\omega_2\sim 1/\tau\lesssim\Lambda_{\text{EFT}}$. The latter demands $\tau T\gtrsim 1/3$ which is satisfied by any $\tau$ obeying the relaxation time bound in strongly coupled systems, namely $\tau T\gtrsim 1$. \footnote{In addition to the conjectured bound mentioned in footnote 1, there is a stronger bound on $\tau$ in a relativistic deformed CFT in 1+1 dimension with finite number of degrees of freedom \cite{Delacretaz:2022ojg}. The latter bound is the consequence of  causality and KPZ universality. }.

In summary,  in the weak coupling limit, which corresponds to $1/(\tau T)\ll1$~\cite{romatschke2019relativistic},   \eqref{QM_to_Classic} reduces to unity and quantum effects are unimportant. In the strong coupling limit,  below $\Lambda_{\text{EFT}}\lesssim 2\pi T$, we can safely use \eqref{QM_to_Classic} to include the quantum effects. Nonetheless, our calculations show that even  the first two sub-leading terms do not contribute significantly to the expression of correlation function. For this reason, in the main text we took $Q=1$ and represented the results for this case.

\section{Microscopic fluctuations}
\label{Appendix_noise}
In the traditional language~\cite{landau2013statistical}, the correlation function \eqref{correlation} is modeled as originating from some microscopic random (noisy) currents, say $\textbf{j}$. Then the noisy dynamics is governed by
\begin{equation}\label{noisy}
	\partial_t n +\boldsymbol{\nabla}\cdot \textbf{J}=\,0\,,\,\,\,\,\,\,\,\,\,\tau\, \partial_t \textbf{J}+\textbf{J}+ D \,\boldsymbol{\nabla}n=\,\textbf{j} \, .
\end{equation}
In this work we assume that these microscopic currents have Gaussian short-distance correlations \footnote{In general, the noise field may have non-vanishing correlation with $n$ (and $\textbf{J}$). This is what is systematically discussed in the framework of Schwinger-Keldysh EFT	\cite{Crossley:2015evo}.  }:
$\langle \text{j}_k(t,\textbf{x})\rangle=\,0\,,\,\,\langle \text{j}_k(t,\textbf{x})\,\text{j}_{\ell}(t',\textbf{x}')\rangle=\,C\,\delta_{j \ell}\,\delta(t-t')\delta(\textbf{x}-\textbf{x}')$.
To determine the strength of the noise, we combine equations \eqref{noisy}:
\begin{equation}\label{noise_theta}
	\tau \partial_t^2  n +\,\partial_t  n-\,D \boldsymbol{\nabla}^2  n=\,\theta \, .
\end{equation}
Here $\theta=- \boldsymbol{\nabla}\cdot\textbf{j}$ is a Gaussian noise field with 
\begin{equation}\label{thetatheta}
	\langle \theta(t,\textbf{x})\,\theta(t',\textbf{x}')\rangle=\,-C(\partial_t)\,\delta(t-t')\boldsymbol{\nabla}^2\delta(\textbf{x}-\textbf{x}') \, .
\end{equation}
Now, from the equation \eqref{noise_theta}, we can simply find $\big(\delta \theta^2\big)_{\omega\textbf{k}}$ in terms of $\big(\delta n^2\big)_{\omega\textbf{k}}$
\begin{equation}\label{nn_thetatheta}
	\langle  n_{\omega \textbf{k}} n_{\omega' \textbf{k}'}\rangle=\,\frac{\langle \theta_{\omega \textbf{k}}\theta_{\omega' \textbf{k}'} \rangle}{(-\tau \omega^2-i \omega+D \textbf{k}^2)(-\tau \omega'^2-i \omega'+D \textbf{k}'^2)} \, ,
\end{equation}
then by using \eqref{correlation}, and considering \eqref{thetatheta},  we arrive at 
\begin{equation}\label{noise_correlatin}
C(\partial_t)\equiv	Q(\partial_t)=\,2\,T  \sigma  \left(\frac{i \partial_t}{2T}\right)\coth\left(\frac{i \partial_t}{2T}\right)\, .
\end{equation}
with $\sigma=\chi D$.
Having specified the correlation function of the noise field, we now want to move on to  constructing the effective action that reproduces all the above correlation functions. We will follow the method developed in 	\cite{Kovtun:2012rj} \footnote{The Schwinger-Keldysh analogue of 	\cite{Kovtun:2012rj} can be found in \cite{Chen-Lin:2018kfl} and \cite{Abbasi:2021fcz}.}.
\section{The effective action}
Equation \eqref{nn_thetatheta} tells us that the average in \eqref{correlation} is in fact over the noise distribution. Then it is reasonable to define the noisy fields, $n_{\theta}(t,\textbf{x})  $, with the subscript $\theta$ denoting that $ n_{\theta}$ is the solution to the equation \eqref{noise_theta}. Taking $W$ as the weight of the noise field distribution, we write 
\begin{equation}\label{Jacobian}
	\begin{split}
		\langle  n(t_1,\textbf{x}_1)   n(t_2,\textbf{x}_2)\rangle=&\,\int \mathcal{D}\theta\, e^{-W[\theta]}\,  n_{\theta}(t_1,\textbf{x}_1)   n_{\theta}(t_2,\textbf{x}_2) \\
		=&\int\mathcal{D}n\int \mathcal{D}\theta\, e^{-W[\theta]}\,\delta(\text{e.o.m.})\,J\, n(t_1,\textbf{x}_1)   n(t_2,\textbf{x}_2) \, ,
	\end{split}
\end{equation}
where  $J=\delta(\text{e.o.m.})/\delta n$ is the Jacobian. The idea of finding the effective action is to exponentiate the above delta function and then integrate over the noise field $\theta$.

At the linearized level, i.e., when e.o.m. is linear, Jacobian is not field-dependent.
By introducing an auxiliary field $n_{a}$, we exponentiate the delta function as
\begin{equation}\label{}
	\langle  n(t_1,\textbf{x}_1)   n(t_2,\textbf{x}_2)\rangle=\int\mathcal{D} n\int \mathcal{D}\theta\, e^{-\frac{1}{2}\int \theta A\theta}\,\int \mathcal{D}n_a e^{i \int (\text{e.o.m.} )n_a}\,J\, n(t_1,\textbf{x}_1)   n(t_2,\textbf{x}_2) \, .
\end{equation}
Considering  \eqref{noise_theta}, the integrating over $\theta$ gives
\begin{equation}\label{}
	\langle  n(t_1,\textbf{x}_1)   n(t_2,\textbf{x}_2)\rangle=\int\mathcal{D}n\, \mathcal{D}n_a\,\, e^{i S_{\text{eff}}^{(2)}[n,\,n_a]}\, n(t_1,\textbf{x}_1)   n(t_2,\textbf{x}_2)\, ,
\end{equation}
where 
\begin{equation}\label{}
	S^{(2)}_{\text{eff}}[n,\,n_a]=\,\int dt d^dx\bigg[i\,n_a \,A\,n_a-\, n_a\left(\tau \partial_t^2n+\partial_t n-D \boldsymbol{\nabla}^2 n\right)\bigg] \, ,
\end{equation}
with  $A=\,-\,T \sigma \left(\frac{i \partial_t}{2T}\right)\coth\left(\frac{i \partial_t}{2T}\right)\boldsymbol{\nabla}^2$. 
From the above quadratic effective action we can read the quadratic Lagrangian; the free propagators then read
\begin{equation}\label{}
	G^{(0)}_{nn_a}=\frac{1}{\omega+ i D \textbf{k}^2- i \tau \omega^2}\,,\,\,\,\,\,\,\,\,\,\,\,G^{(0)}_{n_an}=\frac{-1}{\omega- i D \textbf{k}^2+ i \tau \omega^2} \, ,
\end{equation}
and $G^{(0)}_{nn}$ is given by 
\begin{equation}\label{G_nn_0}
	G^{(0)}_{nn}=\frac{\left(\frac{\omega}{2T}\right)\coth\left(\frac{\omega}{2T}\right)2\,T \chi D \textbf{k}^2}{(\tau \omega^2- D \textbf{k}^2)^2+\omega^2}\,,
\end{equation}
in complete agreement with \eqref{correlation}. Although $G_{nn}$ in this model is well-known from 	Kadanoff-Martin work \cite{kadanoff1963hydrodynamic}, however, \eqref{G_nn_0} is the first derivation of this result from an effective action (involving quantum effects).

Let us consider the non-linear equation of motion~(5). 
Repeating the MSR procedure,  
the Jacobian in \eqref{Jacobian} is no longer field-independent in this case; however, we can exponentiate it by introducing the anticommuting ghost fields $\psi$ and $\bar{\psi}$:
\begin{equation}
	J=\,\int \mathcal{D}\bar{\psi} \mathcal{D}\psi \,e^{-S_{\text{ghost}}}\,,\,\,\,\,\,\,\,\,S_{\text{ghost}}=\,\int d^4x\,\bar{\psi}\frac{\delta \text{E}[n]}{\delta n}\psi \, .
\end{equation}
Then the interacting part of the Lagrangian, up to quartic order, is given by~(6). 
Although we assume that the noise is Gaussian (see \eqref{thetatheta}), but since we expand its magnitude in terms of $n$, i.e. $\sigma(n)=\sigma+\lambda_{\sigma}\delta n+\frac{1}{2}\lambda_{\sigma}'\delta n^2$, we are able to produce the interaction terms $n n_a^2$ and $n^2n_a^2$, consistent with the result of Schwinger-Keldysh framework \cite{Chen-Lin:2018kfl}. We would also like to point out that we did not apply any constraints from the fluctuation-dissipation theorem to the interaction terms in the action. This is indeed beyond the scope of the MSR formalism. Here we only apply the fluctuation dissipation theorem to the two-point functions in quadraic part of the cation. A detailed analysis of KMS constraints for higher n-point functions, or the so-called generalized fluctuation dissipation theorem, can be found in \cite{Wang:1998wg}. See also \cite{Chao:2020kcf} for an interesting discussion of $\mathcal{T}-$reversal symmetry in a system with self-interacting stochastic fields. This symmetry is closely related to the KMS symmetry.

As the last point, note that, as discussed in \cite{Kovtun:2012rj} the ghost terms do not contribute to the following computations.

\section{Loop calculations}
\label{Appendix_loop_calculations}
In order to study the effect of hydrodynamic interactions on the correlation functions, we parameterize the 1-loop corrections to $G_{n n_a}$ as the following ($p=(\omega, \textbf{k})$):
\begin{equation}\label{Sigma}
	G^{(1)}_{n n_a}(p)=\,G^{(0)}_{n n_a}(p)+G^{(0)}_{n n_a}(p)(-\Sigma(p))G^{(0)}_{n n_a}(p)
	=\,\frac{1}{\omega+ i D_0 \textbf{k}^2- i \tau\, \omega^2+\,\Sigma(\omega, \textbf{k})} \, ,
\end{equation}
with $\Sigma$ being the self-energy appearing in the retarded Green's function~(7). 
It turns out that there are only two one-loop diagrams contributing to the self-energy 
\begin{equation}\label{Sigma_0}
	\begin{split}
		G^{(0)}_{n n_a}\Sigma(p)G^{(0)}_{n n_a}=&\,
		\scalebox{0.53}{	\begin{tikzpicture}[baseline=(a.base)]
				\begin{feynman}
					\vertex (a) ;
					\vertex [right=of a] (a1) ;
					\vertex [right=of a1] (a2);
					\vertex [above right=of a2] (a3);
					\vertex [below right=of a2] (a4);
					\vertex [above right=of a4] (a5); 
					\vertex [ right=of a5] (a6);
					\vertex [ right=of a6] (a7);
					\diagram* {
						(a) --[very thick](a1)-- [boson,very thick] (a2) -- [quarter left, very thick] (a3)--[boson,quarter left,very thick](a5)--[quarter left, very thick](a4)--[quarter left,very  thick](a2) ,
						(a5) -- [very thick] (a6)--  [boson,very thick](a7),
					};
				\end{feynman}
		\end{tikzpicture}}
		+
		\scalebox{0.53}{	\begin{tikzpicture}[baseline=(a.base)]
				\begin{feynman}
					\vertex (a) ;
					\vertex [right=of a] (a1) ;
					\vertex [right=of a1] (a2);
					\vertex [above right=of a2] (a3);
					\vertex [below right=of a2] (a4);
					\vertex [above right=of a4] (a5); 
					\vertex [ right=of a5] (a6);
					\vertex [ right=of a6] (a7);
					\diagram* {
						(a) --[very thick](a1)-- [boson,very thick] (a2) -- [quarter left, very thick] (a3)--[quarter left,boson,very thick](a5)--[quarter left,boson,very thick](a4)--[quarter left, very thick](a2) ,
						(a5) --  [very thick](a6)--  [boson,very thick](a7),
					};\,.
				\end{feynman}
		\end{tikzpicture}} \, .
	\end{split}
\end{equation}
In fact, $\Sigma$ may correct $D$ and $\tau$, as we will discuss below.
Then we find
\begin{equation}\label{}
	\begin{split}
		\Sigma(p)=\,&\,\lambda_D^2\,\textbf{k}^2\int_{p'}\textbf{k}'^2G^{(0)}_{n_a n}(p')G^{(0)}_{n n}(p'+p)\\
		\,\,&-\frac{i}{2} \chi T\lambda_D \lambda_{\sigma} \textbf{k}^2\int_{p'}\,\,(\textbf{k}'^2+ \textbf{k}\cdot\textbf{k}')\bigg[Q(\omega')+Q(\omega+\omega')\bigg]G^{(0)}_{n_a n}(p')G^{(0)}_{n n_a}(p+p') \, ,
	\end{split}
\end{equation}
with $Q(\omega)$ defined in \eqref{QM_to_Classic}.
In order to evaluate the frequency integrals, we need to know the analytic structure of the integrand. Each of Green's functions has only two or four simple poles. The expression  $Q(\omega')$, however,   has infinite number of poles on the imaginary axis in the complex frequency plane. As discussed earlier, as long as the system under study satisfies $\Lambda_{\text{EFT}}\lesssim 2\pi T$, we can safely expand it about $\omega'=0$.     \textit{Let us emphasize that in this way we will include all quantum effects in addition to statistical fluctuations.}

Taking a hard momentum cutoff, performing standard although lengthy calculations, the cutoff-independent part of $\Sigma$ is found to be given by~(9) 
and~(10). 
At $d=1$ the functions $f_1$ and $f_2$ in~(9) 
are given by
\begin{equation*}
	\begin{split}
		f_1(\omega,\textbf{k})&= \,\frac{ \omega (1-i\tau \omega)(D \textbf{k}^2 \tau -\tau ^2 \omega ^2-3 i \tau  \omega +2)^2}{4\big(D \textbf{k}^2 \tau +(1-i\tau  \omega)^2\big)^2}\left[1+\frac{1}{(T \tau)^2}Q_{1}^{(2)}+\frac{1}{(T \tau)^4}Q_{1}^{(4)}\right] \, , \\
		Q_{1}^{(2)}&=-\frac{(D \textbf{k}^2\tau-i \tau \omega(1-i \tau \omega))^2}{48(D \textbf{k}^2\tau+(1-i \tau \omega)^2)} \, , \\
		Q_{1}^{(4)}&=\frac{(D \textbf{k}^2\tau-i \tau \omega(1-i \tau \omega))^2\big(3 (D \textbf{k}^2\tau)^2-2 (i \tau\omega) (D \textbf{k}^2\tau)(3-i \tau \omega)-(i \tau\omega)^2(1- i \tau \omega)^2\big)}{11520(D \textbf{k}^2\tau+(1-i \tau \omega)^2)^2} \, , 
	\end{split}
\end{equation*}
and 
\begin{equation*}
	\begin{split}
		f_2(\omega,\textbf{k})&=-\frac{ i (D \textbf{k}^2 - i \omega - i \tau \omega^2)   (1-i\tau \omega)(D \textbf{k}^2 \tau -\tau ^2 \omega ^2-3 i \tau  \omega +2)}{2\big(D \textbf{k}^2 \tau +(1-i\tau  \omega)^2\big)^2}\left[1+\frac{1}{(T \tau)^2}Q_{2}^{(2)}+\frac{1}{(T \tau)^4}Q_{2}^{(4)}\right] \, , \\
		Q_{2}^{(2)}&=\frac{(D \textbf{k}^2\tau)^2-2 (i\tau \omega)(D \textbf{k}^2\tau)-(i \tau\omega)^2(1- i \tau \omega)^2}{48(D \textbf{k}^2\tau+(1-i \tau \omega)^2)}\, , \\
		Q_{2}^{(4)}&=-\frac{1}{11520(D \textbf{k}^2\tau+(1-i \tau \omega)^2)^2}\bigg[(D \textbf{k}^2\tau)^4-4 (D \textbf{k}^2\tau)^3(i \tau \omega)(1+i \tau \omega) \, , \\
		&\,-2(D \textbf{k}^2\tau)^2(i \tau \omega)^2(1-10i \tau \omega-5 \tau^2\omega^2)+4(D \textbf{k}^2\tau)(i \tau\omega)^3(1-i \tau \omega)^2(3- i \tau \omega)+(i \tau\omega)^4(1- i\tau \omega)^4\bigg]\, .
	\end{split}
\end{equation*}
We do not represent the structure of $f_1$ and $f_2$  at $d=2,3$ here. As was also mentioned earlier, $Q_{2}^{(2)}$ and $Q_{2}^{(4)}$ are suppressed by smaller numerical factors compared to the leading expression.  

So far we have found $\Sigma$ 
in~(7). 
It is easy to show that the pole singularities of the retarded Green's function~(7) 
are the roots of the following equation\footnote{We thank Pavel Kovtun for discussion on this point.}
\begin{equation}
	(\omega+i D \textbf{k}^2-i \tau \omega^2)\bigg(1-\frac{\delta\sigma(\omega,\textbf{k})}{\sigma}\bigg)+\Sigma(\omega,\textbf{k})=\,0 \, .
\end{equation}
To find $\delta\sigma(p)$, we apply the FDT theorem to~(7) 
and first obtain $G_{nn}$ as follows
\begin{equation}\label{G_{nn}}
	G^{(1)}_{nn}(\omega,\textbf{k})=\,\frac{N(\omega,\textbf{k})}{\omega^2+D^2 \textbf{k}^4+2\,\omega\,\text{Re}\,\Sigma(\omega,\textbf{k})+2\,(D \,\textbf{k}^2- \tau \omega^2)\,\text{Im}\,\Sigma(\omega,\textbf{k})}\, .
\end{equation}
Here the numerator $N(p)$ contains $\delta{\sigma}(p)$ as follows 
\begin{equation}\label{C}
	N(p)=2 T \chi D  \textbf{k}^2Q(\omega)\left[1+\frac{\text{Re}\delta \sigma(p)}{\sigma}+\frac{D \textbf{k}^2-\tau \omega^2}{\omega}\frac{\text
		{Im}\delta \sigma(p)}{\sigma}+\frac{\text{Re}\Sigma(p)}{\omega}\right] \, .
\end{equation}
We then need to calculate $N(p)$. Diagrammatically, we have
\begin{equation*}
	\begin{split}
		&G^{(0)}_{n n_a}(p)(-N(p))G^{(0)}_{n_a n}(p)=\\
		&\feynmandiagram[scale=0.5,transform shape] [inline=(a.base),horizontal=a to b] {a --[thick]c-- [thick]b};
		+
		\scalebox{0.65}{	\begin{tikzpicture}[baseline=(a.base)]
				\begin{feynman}
					\vertex (a) ;
					\vertex [right=of a] (a1) ;
					\vertex [right=of a1] (a2);
					\vertex [above=of a2] (a3);
					\vertex [ right=of a2] (a4);
					\vertex [ right=of a4] (a5); 
					\diagram* {
						(a) -- [ thick](a1)-- [boson,very thick] (a2) -- [out=140, in=180, loop, min distance=1.cm, thick] (a3)--[out=0, in=40, loop, min distance=1.cm, thick](a2) ,
						(a2) -- [boson,very thick] (a4)--   [ thick](a5),
					};
				\end{feynman}
		\end{tikzpicture}}
		+
		\scalebox{0.5}{	\begin{tikzpicture}[baseline=(a.base)]
				\begin{feynman}
					\vertex (a) ;
					\vertex [right=of a] (a1) ;
					\vertex [right=of a1] (a2);
					\vertex [above right=of a2] (a3);
					\vertex [below right=of a2] (a4);
					\vertex [above right=of a4] (a5); 
					\vertex [ right=of a5] (a6);
					\vertex [ right=of a6] (a7);
					\diagram* {
						(a) -- [ very thick](a1)-- [boson,very thick] (a2) -- [quarter left, very thick] (a3)--[quarter left, very thick](a5)--[quarter left, very thick](a4)--[quarter left, very thick](a2) ,
						(a5) -- [boson,very thick] (a6)--   [very thick](a7),
					};
				\end{feynman}
		\end{tikzpicture}}
		+
		\left [\, \scalebox{0.5}{	\begin{tikzpicture}[baseline=(a.base)]
				\begin{feynman}
					\vertex (a) ;
					\vertex [right=of a] (a1) ;
					\vertex [right=of a1] (a2);
					\vertex [above right=of a2] (a3);
					\vertex [below right=of a2] (a4);
					\vertex [above right=of a4] (a5); 
					\vertex [ right=of a5] (a6);
					\vertex [ right=of a6] (a7);
					\diagram* {
						(a) --[very thick](a1)-- [boson,very thick] (a2) -- [quarter left, very thick] (a3)--[quarter left,boson,very thick](a5)--[quarter left, very thick](a4)--[quarter left, very thick](a2) ,
						(a5) -- [boson,very thick] (a6)--  [very thick](a7),
					};
				\end{feynman}
		\end{tikzpicture}}
		+\text{c.c} \right]\,\, ,
	\end{split}
\end{equation*}
which leads to
\begin{equation}\label{C_int}
	\begin{split}
		N(p)=\,&\,2 T \chi D  \textbf{k}^2 Q(\omega)+ 2 \chi T \lambda'_{\sigma} Q(\omega)\int_{p'}G^{(0)}_{nn}(p')+
		\frac{1}{2}\lambda_D^2\textbf{k}^4\int_{p'}G^{(0)}_{nn}(p')G^{(0)}_{nn}(p-p')\\
		+\,&i \chi T \lambda_{\sigma}\lambda_D \textbf{k}^2
		\int_{p'}\textbf{k}\cdot \textbf{k}'G^{(0)}_{n_a n}(p')G^{(0)}_{nn}(p+p')(Q(\omega)+Q(\omega')) \, .
	\end{split}
\end{equation}
Once we find $N(p)$ in \eqref{C_int}, we can read $\delta\sigma(p)$ from \eqref{C}.  It turns out that to leading order in the expansion of $C$ in \eqref{QM_to_Classic}, $f_{3}$ vanishes. The expression of $f_{4}$ for $d=1$ is given by: 
\begin{equation}
		f_{4}(\omega,\textbf{k})= \,-\frac{ (1-i\tau \omega)(D \textbf{k}^2 \tau -\tau ^2 \omega ^2-3 i \tau  \omega +2)}{2\big(D \textbf{k}^2 \tau +(1-i\tau  \omega)^2\big)^2}\,.
\end{equation}
Theoretically, it would be interesting to realize the results of this letter in the framework of Schwinger-Keldysh EFT~\cite{Crossley:2015evo,Jensen:2018hse,Haehl:2015pja}. To our knowledge, inclusion of gapped modes beyond quadratic order~\cite{Baggioli:2020haa,Delacretaz:2021qqu} has not been explored in this context yet.

\section{Threshold singularities}
\label{}
Considering the branch points given in Eq.~(14), $\tilde{\omega}_{11}$ and $\tilde{\omega}_{22}$ correspond to the minimum energy to generate a pair of on-shell $\omega_1$ and $\omega_2$ excitations in the loop, respectively. 
For instance, in the case of $\tilde{\omega}_{11}$, we can simply consider the situation in which the two lines of the second loop in~(8) 
carry the excitation $\omega_{1}(\textbf{k})$. Conservation of energy and momentum then enforces 
$	\omega(\textbf{k})=\,\omega_1(\textbf{k}')+\omega_1(\textbf{k}-\textbf{k}')\,.$ 
Using~(3), 
we find that $\omega(\textbf{k})$ becomes the minimum value at $\textbf{k}'=\textbf{k}/2$, and this value is equal to $\tilde{\omega}_{11}$ in~(14). 

\section{Dispersion relations}
\label{}

\begin{figure}[htbp]
	\centering
	\includegraphics[width=0.47\textwidth]{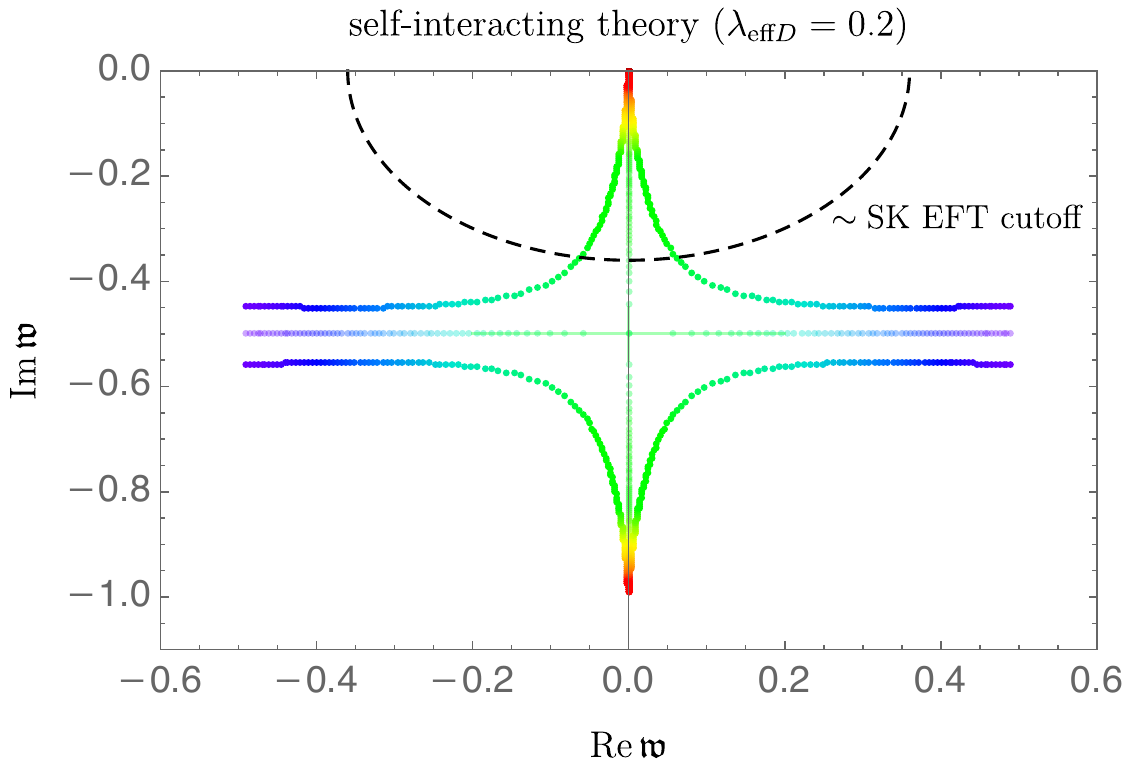}
	\caption{Spectrum of the excitations in the complex $\wn=\omega \tau$ plane. 
		Each colored trajectory, starting with dark red and ending with purple, shows the change of a particular mode when $\qn=(D \textbf{k}^2 \tau)^{1/2}$  increases from $0$ to $0.8$. 
		The two modes of the non-interacting theory given by~(3) 
		are shown in low opacity. The  dashed semicircle  is schematically showing the cutoff of  Schwinger-Keldysh EFT of diffusive fluctuations discussed in Refs.~\cite{Chen-Lin:2018kfl,Abbasi:2021fcz}.
	}
	\label{modes_d_1}
\end{figure}

From~(12), 
we can also extract the dispersion relation of fluctuations.
Defining the dimensionless effective coupling constant and momentum as  $\lambda_{\text{eff}D} =\frac{1}{16D}\frac{(T \chi)^{1/2}}{(D \tau)^{d/4}}\lambda_D$, we have numerically illustrated the spectrum of fluctuations in Fig.~\ref{modes_d_1} for $d=1$. For $d=2,3$, the spectrum is qualitatively similar.
We find that due to the self-interactions, \textit{first}, each of the two modes $\omega_1$ and $\omega_2$ splits into two modes. 
\textit{Second}, the split dispersion relations avoid a level-crossing; the stronger the self-interactions, the more repulsion between them. 

\section{Long-time-tails}
\label{long_time_tail}
In this section we find the long time tail behavior of $\mathfrak{G}^{R(1)}_{nn}$ and $\mathfrak{G}^{(1)}_{nn}$. Let us start by expanding~(12). 
We have
\begin{equation}   \label{} 
	\begin{split} 
		G^{R(1)}_{nn}(\omega\,, \textbf{k})=\frac{\sigma\,\textbf{k}^2}{-  \tau \omega^2- i \omega +D \textbf{k}^2 }-\frac{\sigma \,\textbf{k}^4\delta D(\omega, \textbf{k})}{(-  \tau \omega^2- i \omega +D \textbf{k}^2)^2}
	\end{split}
\end{equation}
Now by using the fluctuation dissipation theorem, $G_{nn}= \frac{2T}{\omega} \text{Im} G^{R}_{nn}$, we find
\begin{equation}   \label{G_nn_1} 
	\begin{split} 
		G^{(1)}_{nn}(\omega\,, \textbf{k})=\frac{2 T \chi D \,\textbf{k}^2}{\omega^2+(\tau \omega^2 -D \textbf{k}^2)^2}-\frac{2}{\omega}\text{Im}\frac{T \chi D \,\textbf{k}^4\delta D(\omega, \textbf{k})}{(-  \tau \omega^2- i \omega +D \textbf{k}^2)^2}
	\end{split}
\end{equation}
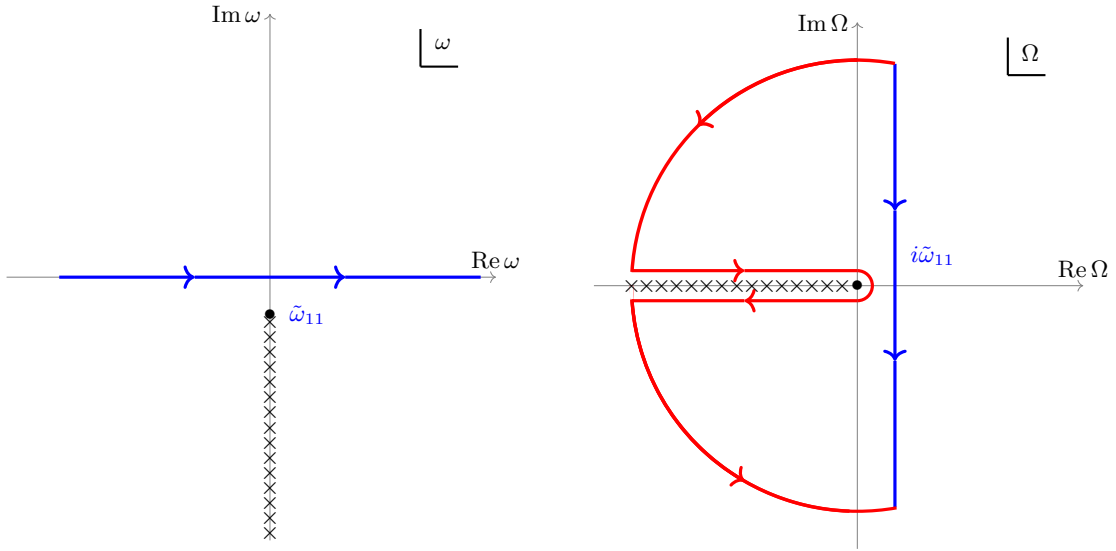
\begin{figure}
	\centering
	\begin{tikzpicture}
		[decoration={markings,
		}
		]
		\draw[help lines,->] (-3.5,0) -- (3.,0) coordinate (xaxis);
		\draw[help lines,->] (0,-3.5) -- (0,3.5) coordinate (yaxis);
		\draw[draw,blue,line width=1.2pt,->] (-2.8,0) -- (-1,0) coordinate;
		\draw[draw,blue,line width=1.2pt,->] (-1,0) -- (1,0) coordinate;
		\draw[draw,blue,line width=1.2pt] (1,0) -- (2.8,0) coordinate;
		\draw[draw,black,line width=.8pt] (2,2.8) -- (2.5,2.8) coordinate;
		\draw[draw,black,line width=.8pt] (2,2.8) -- (2,3.3) coordinate;
		\node[] at (0,-3.4) {$\times$};
		\node[] at (0,-3.2) {$\times$};
		\node[] at (0,-3) {$\times$};
		\node[] at (0,-2.8) {$\times$};
		\node[] at (0,-2.6) {$\times$};
		\node[] at (0,-2.4) {$\times$};
		\node[] at (0,-2.2) {$\times$};
		\node[] at (0,-2.) {$\times$};
		\node[] at (0,-1.8) {$\times$};	
		\node[] at (0,-1.6) {$\times$};	
		\node[] at (0,-1.4) {$\times$};
		\node[] at (0,-1.2) {$\times$};
		\node[] at (0,-1.) {$\times$};
		\node[] at (0,-.8) {$\times$};	
		\node[] at (0,-.6) {$\times$};			
		\node[] at (0,-.5) {$\bullet$};
		\node[above] at (xaxis) {$\text{Re}\, \omega$};
		\node[left] at (yaxis) {$\text{Im} \,\omega$};
		\node[blue] at (.5,-.5) {$\tilde{\omega}_{11}$};
		\node[black] at (2.3,3.1) {$\omega$};
	\end{tikzpicture}
	\,\,\,\,\,\,\,\,\,\,\,\,
	\begin{tikzpicture}
		[decoration={markings,
		}
		]
		\draw[help lines,->] (-3.5,0) -- (3.,0) coordinate (xaxis);
		\draw[help lines,->] (0,-3.5) -- (0,3.5) coordinate (yaxis);
		\draw[draw,blue,line width=1.2pt,->] (.5,2.95) -- (.5,1) coordinate;
		\draw[draw,blue,line width=1.2pt,->] (.5,1) -- (.5,-1) coordinate;
		\draw[draw,blue,line width=1.2pt] (.5,-1) -- (.5,-2.95) coordinate;
		\draw[draw,black,line width=.8pt] (2,2.8) -- (2.5,2.8) coordinate;
		\draw[draw,black,line width=.8pt] (2,2.8) -- (2,3.3) coordinate;
		\path[draw,line width=1.3pt,red,postaction=decorate] (.5,2.955) arc (80:90.5:3);
		\path[draw,line width=1.3pt,red,postaction=decorate,->] (0,3) arc (90:135:3);
		\path[draw,line width=1.3pt,red,postaction=decorate] (0,3) arc (90:176.5:3);
		\path[draw,line width=1.3pt,red,postaction=decorate] (-3,0) arc (180:268:3);
		\path[draw,line width=1.2pt,red,postaction=decorate,->] (-3,0) arc (180:240:3);
		\path[draw,line width=1.2pt,red,postaction=decorate] (-3,0) arc (180:280:3);
		\path[draw,line width=1.4pt,white,postaction=decorate] (-3,0) arc (180:183.5:3);
												
		\draw[draw,red,line width=1.2pt,->] (0,-.2) -- (-1.5,-.2) coordinate;
		\draw[draw,red,line width=1.2pt] (-3,-.2) -- (-1.5,-.2) coordinate;
		\draw[draw,red,line width=1.2pt] (0,.2) -- (-1.5,.2) coordinate;
		\draw[draw,red,line width=1.2pt,->] (-3,.2) -- (-1.5,.2) coordinate;
		\path[draw,line width=1.2pt,red,postaction=decorate] (0,.2) arc (90:-90:.2);

		\node[] at (-3,0) {$\times$};
		\node[] at (-2.8,0) {$\times$};
		\node[] at (-2.6,0) {$\times$};
		\node[] at (-2.4,0) {$\times$};
		\node[] at (-2.2,0) {$\times$};
		\node[] at (-2.,0) {$\times$};
		\node[] at (-1.8,0) {$\times$};	
		\node[] at (-1.6,0) {$\times$};	
		\node[] at (-1.4,0) {$\times$};
		\node[] at (-1.2,0) {$\times$};
		\node[] at (-1.,0) {$\times$};
		\node[] at (-.8,0) {$\times$};	
		\node[] at (-.6,0) {$\times$};	
		\node[] at (-.4,0) {$\times$};
		\node[] at (-.2,0) {$\times$};	
		\node[] at (0,0) {$\bullet$};
		\node[above] at (xaxis) {$\text{Re}\, \Omega$};
		\node[left] at (yaxis) {$\text{Im} \,\Omega$};
		\node[blue] at (1,.4) {$i \tilde{\omega}_{11}$};
		\node[black] at (2.3,3.1) {$\Omega$};
	\end{tikzpicture}
\caption{ Left panel:  The dominant part of the analytic structure of $G_{nn}(\omega, \textbf{k})$ at late times ($t>>t_{D}$). Right pane: Changing the integral variable from $\omega$ to $\bar{\omega}$ and contour deformation. Note that $i \omega_{11}=1-\sqrt{1- \tau D \textbf{k}^2}$ is real-valued.}
\label{contour_long_time_tail}
\end{figure}
%
To perform the Fourier integral  $\mathfrak{G}_{nn}(t,\textbf{k})=\int \frac{d\omega}{2\pi}G_{nn}(\omega,\textbf{k})\,e^{-i \omega t}$ at $t>0$ we consider the analytic structure of  $G_{nn}(\omega,\textbf{k})$ in the lower half $\omega$ plane. It is shown in Fig. 1. 
In the asymptotic limit
\begin{equation}   \label{} 
	\tau\ll \frac{1}{D \textbf{k}^2}\lesssim t
\end{equation}
 the dominant contribution comes from  the branch point closest to the real axis, namely $\tilde{\omega}_{11}$.  Thus to find the long time tail, or equivalently the asymptotic behavior of $\mathfrak{G}_{nn}(t,\textbf{k})$, we need to expand $G_{nn}(\omega,\textbf{k})$ about $\tilde{\omega}_{11}$ and keep only the leading term. For this leading term, the branch cut structure is given in the left panel of Fig.~\ref{contour_long_time_tail}. This is nothing but the the region surrounded by the dashed semicircle in figure 1. 
 
  Now taking $\omega=\tilde{\omega}_{11}+i\Omega$, we can deform the contour of integration, as depicted in te right panel of Fig.~\ref{contour_long_time_tail}. The Fourier integral takes the following form (at $d=1$)
\begin{equation}   \label{} 
	\mathfrak{G}_{nn}(t,\textbf{k})=g\frac{\big(1+\sqrt{1- \tau D \textbf{k}^2}\big)^4}{(1- \tau D \textbf{k}^2)^{1/4}}\,\frac{e^{-i \tilde{\omega}_{11} t}}{\sqrt{2D}}\int_{-\infty}^{0} \frac{i d\Omega}{2\pi}\,\text{Disc}\frac{e^{ \Omega t}}{\sqrt{\Omega}}
\end{equation}
where $g=\frac{\lambda_{D}^2}{16D^2}T^2 \chi^2$ and $\text{Disc} f(z)=\lim\limits_{\epsilon\to 0}f(z+ i \epsilon)-f(z-i \epsilon)$.
The integral in the above expression evaluates to $1/\sqrt{\pi t}$ and we find the result given in~(15).  

\section{The effect of UV regulator on fluctuations in expanding QGP}
\label{QGP}
Let us recall that the linear response of the system to external sources can be found through the linear response framework. For the system under the consideration in this letter, 
we may consider a chemical potential source at $t<0$ that is turned off at $t=0$: $\mu(t,\textbf{x})=e^{\epsilon t}\mu(\textbf{x})\theta(-t)$ \cite{Kovtun:2012rj}. Then 
\begin{equation}   \label{} 
\langle n(t, \textbf{x}) \rangle=\int_{\infty}^0 dt'e^{\epsilon t'}\int d^4x' \mu(\textbf{x}')G^{R(0)}_{nn}(t-t',\textbf{x}-\textbf{x}')
\end{equation}
Note that $G^{R(0)}_{nn}$ is the regarded Green's function in the absence of non-linear interactions. 
We find
\begin{equation}   \label{n_t_k} 
	\langle n(t, \textbf{k}) \rangle=\mu_0(\textbf{k})\int_{\infty}^0 \frac{d \omega}{2\pi}G^{(0)R}_{nn}(\omega, \textbf{k})\frac{e^{- i \omega t }}{i \omega +\epsilon}
\end{equation}
where $\mu_0(\textbf{k})$ is the Fourier transform of $\mu(t,\textbf{x})$ at $t=0$.
We would like to use these expressions for the case of Bjorken flow. Since the flow profile only depends on the proper time $\tau_p$, we need to have a dynamic quantity that has only a time-dependence. For this reason, we choose to introduce $\langle n(t) \rangle\equiv\langle n(t, \textbf{x}=0) \rangle$. It is given by
\begin{equation}   \label{n_t} 
	\langle n(t)\rangle=\int d^3 k 	\langle n(t, \textbf{k}) \rangle
\end{equation}
To find this, two things need to be determined in \eqref{n_t_k}; $	G^{(0)R}_{nn}(\omega,\textbf{k})$ and $\mu_0(\textbf{k})$.
Let us emphasize that our final goal is to find the correction to $	\langle n(t)\rangle$ caused by the non-linear interactions. Let us call it $\Delta 	\langle n(t)\rangle$.  For this, we evaluate the integral in \eqref{n_t_k} when $G^{(0)R}_{nn}(\omega, \textbf{k})$ is replaced with $G^{(1)R}_{nn}(\omega, \textbf{k})$.
The latter is found from our EFT calculations 
\begin{equation}\label{G_R_3}
	G^{(1)R}_{nn}(\omega, \textbf{k})=-\frac{\lambda_D^2 T \chi^2}{128\pi D^{3/2}}
	\frac{(i \omega)\,\textbf{k}^4}{\big(D \text{k}^2-\omega (i+\tau \omega)\big)^2}\left(\frac{2+D \textbf{k}^2\tau-\tau \omega(3i+\tau \omega)}{- D \textbf{k}^2\tau+(i +\tau \omega)^2}\right)^2\sqrt{\frac{D \textbf{k}^2-\omega  (2i+\tau  \omega)}{D \textbf{k}^2 \tau -(i+\tau  \omega)^2}}
\end{equation}
To specify $\mu_{0}(\textbf{k})$ we take it into account that in the presence of an external deriving frequency $\omega_{ext}$, there will be an important length scale in the system, the so-called dissipative scale $k^{*}$
\begin{equation}   \label{} 
D k_*^2\sim \omega_{ext}\,\,\,\,\,\,\, \rightarrow\,\,\,\,\,\,\,\,\, k_*\sim \left(\frac{\omega_{ext}}{D}\right)^{1/2}
\end{equation}
Modes with $k\gg k_*$ have been already equilibrated. Modes with $k\ll k_*$ are out of equilibrium and evolve according to linear hydrodynamics. However, for modes $k\sim k_*$, the equilibrium is ongoing. These modes contribute to the hydrodynamic fluctuations through non-linearities \cite{Akamatsu:2016llw}. To exclusively estimate the contribution of these modes, we take the chemical potential to have the following form   
\begin{equation}   \label{mu_0} 
\mu_0(\textbf{k})= 2\pi^2\,\mu\,\tau_{ext} D\, \delta\bigg(k-\frac{1}{\sqrt{\tau_{ext} D}}\bigg)
\end{equation}
where $\tau_{ext} \sim \omega_{ext}^{-1}$ and $\mu$ is the amplitude of $\mu(t, \textbf{x})$ at $t=0$. The normaliztion factor has been chosen so that $\int \frac{d^3k}{(2\pi)^3}\mu_0(\textbf{k})=\mu$.

Now we have the necessary ingredients, namely \eqref{G_R_3} and  \eqref{mu_0}, to calculate \eqref{n_t}. We perform the calculations in Bjorken flow. In this case, $\omega_{ext}$ is nothing but the expansion rate of flow, i.e.; $1/\tau_{p}$ with $\tau_p$ being the proper time. The last step is to identify $t$ with $\tau_{p}$ and to evaluate the integral in \eqref{n_t}. The result is 
\begin{equation}   \label{n_tau_p} 
	\Delta\langle n(\tau_p)\rangle=\bigg(\frac{g\,\mu}{4 T}\bigg)\,(1- \frac{\tau}{\tau_{p}})^{1/4}\bigg(1+\sqrt{1- \frac{\tau}{\tau_{p}}}\bigg)^4\,\frac{e^{\big(-1+\sqrt{1- \frac{\tau}{\tau_{p}}}\big)\frac{\tau_p}{\tau}}}{(2\pi D \tau_p)^{3/2}}\,.
\end{equation}
The timescale set by the UV-regulator, namely $\tau$ is much smaller than the proper time $\tau_p$: $\tau\ll \tau_p$. See equation (12) in \cite{Akamatsu:2016llw}. In our notation, it is expressed as $\frac{1}{c_s \tau_{p}}\ll k_*\ll \frac{1}{c_s \tau}$. Then it is evident that $\tau_p\gg \tau$. 

To gain more insight into \eqref{n_tau_p}, we expand it in powers of $\frac{\tau}{\tau_{p}}$:
\begin{equation}   \label{} 
	\Delta\langle n(\tau_p)\rangle=\frac{1}{4\sqrt{e(2\pi)^3}}\,T \chi^2 \mu\,\frac{\lambda_D^2}{D^2}\,\frac{1}{(D \tau_p)^{3/2}}\bigg(1-\frac{11}{8}\frac{\tau}{\tau_p}+\frac{49}{128}\frac{\tau^2}{\tau_p^2}+\cdots\bigg)\,.
\end{equation}
A more realistic scenario should include not only diffusion. This requires improving the present work by considering the full MIS equations~\cite{israel1979transient,muller1967paradoxon,israel1976nonstationary}, for recent presentations see~\cite{romatschke2019relativistic,jeon2015introduction,gale2013hydrodynamic}. 
It would be interesting to construct the hydro-kinetic equations for this full set of MIS equations. 

\section{Fitting method}
\label{fit_method}
To evaluate the possibility of non-linear effects in the bad metallic system examined in \cite{Brown:2018}, the authors of  \cite{Brown:2018} fit the experimental results to $\delta n(t, \textbf{k } )$ (6 different values of $\textbf{k}$) to the analytical solution of equation~(2). 
The solution is parameterized by three parameters: $\Gamma$, $D$ and the amplitude $\delta n(0, \textbf{k})$.
For any set of the data points associated with a specific value of $\textbf{k}$, they find these parameters. The results are given in  plots S1-B and S1-C of this reference. 
\vspace{2mm}

Our goal is to fit the data with a solution to the non-linear equation~(5) 
($\lambda_D'=0$). However, this equation has no analytical solution. Therefore, we must fit the data to its numerical solution. After trying different fitting \texttt{Method}s in Mathematica, we came to the conclusion that in order to find reliable fitting results,
\begin{enumerate}
		\item Linear and non-linear differential equations must be solved in the same way.
	\item Both linear and non-linear fits must be done using the same method.
	\end{enumerate}
The first item above means that both linear and non-linear equations should be solved numerically. This is why the linear fit (orange and red) points in Figure 3 
do not precisely coincide with  the results from \cite{Brown:2018}. Reference \cite{Brown:2018} chooses to fit the data to an analytical solution of~(2), 
whereas we do this by fitting a numerical solution of~(2). 
We have used
 \begin{verbatim}
NDSolve[... , 	Method-> "StiffnessSwitching"]
	\end{verbatim} 
{The solution of a linear equation at any $\textbf{k}$ is parameterized by three parameters $\Gamma$, $D$ and $A\equiv \delta n(0,\textbf{k})$. However, the solution to a nonlinear equation is specified by four parameters: $\Gamma$, $D$, $A$, and $\lambda$. Therefore we call the solutions of linear and nonlinear equations } 
{\verb| solLinear[GammaL, DL, AL][t]|} and {\verb| solNonLinear[GammaN, DN, AN, lambdaN][t]|}, 
{respectively.} \\
For the second item above, we found that the appropriate method is \texttt{NMinimize}. For the case of linear fitting, we simply do it as follows
\begin{verbatim}
	FindFit[data, solLinear[GammaL, DL, AL][t], {GammaL, DL, AL}, t, Method-> NMinimize]
\end{verbatim} 
At any $\textbf{k}$, the result is the set 
\begin{verbatim}
{GammaL, DL, AL}
\end{verbatim} 
The results are given in Table.~\ref{table_near_horizon}.
\begin{table}[h]
	\centering
	\begin{tabular}{|c|c|c|c|}
		\hline
		$\text{k}$ & GammaL & DL & AL \\
		\hline
		\hline
		0.523599 & 0.00795346 & 0.0261384& 0.138968 \\
		\hline
		0.498666 &0.0104197  &0.0249761 & 0.085029\\
		\hline
		0.541654 &0.0100459 &0.0186697 & 0.088081\\
		\hline
		0.515015 & 0.00513258& 0.0347175&0.0543277 \\
		\hline
		0.546364 & 0.00925405& 0.0216909&0.054187 \\
		\hline
		0.537024 & 0.00476434&0.0383528 &0.0353531 \\
		\hline
		0.610018 & 0.00540354&0.0284217 &0.0342472 \\
		\hline
		0.506708 & 0.00821242& 0.0254027& 0.103702\\
		\hline
	\end{tabular}
	\caption{Linear fitting results with the NMinimize method.} 
	\label{table_near_horizon}
\end{table}
%

However, for the case of fitting data to the non-linear equation, the situation is more complicated. We were unable to find any non-linear smooth fit to the original data. Instead, we find that at any $\textbf{k}$ it is better to use the solution of the linear equation, i.e. \texttt{solLinear[GammaL, DL, AL][t]} as the benchmark. We then generate a new set of data from it. The number of new data points read from this function depends on $\textbf{k}$. We define
\begin{verbatim}
	Data[M_]:=Table[solLinear[GammaL, DL, AL][j M], {j,1,N}]
\end{verbatim}
\so{Among them, the maximum value of \texttt{N} is found by requiring \texttt{N M} not to exceed the number of data points. Then at any $\textbf{k}$ we ask for the output} 
Considering the function \verb|solLinear[GammaL, DL, AL][t]| 
at a specific value of $\textbf{k}$, the above command simply selects the value of the function  at successive instants 
$ t=\verb|0|, \verb|M|,  \verb|2|\,\verb|M|, \cdots, \verb|N|\,\verb|M| $. 
{Two things must be determined here: } \verb|M| and \verb|N|. 
{The value of \texttt{N} is found by requiring \texttt{N M} not to exceed the time associated with the data point at the largest $t$. Then at any $\textbf{k}$ we ask for the output of }
\begin{verbatim}
	NonLinearModelDFit[data[M], solNonLinear[GammaN, DN, AN, lambdaN][t],
	{GammaN, DN, AN, lambdaN}, t, method->NMinimize]
\end{verbatim}
to satisfy
\begin{verbatim}
	Abs[solNonLinear[GammaN, DN, AN, lambdaN][t]-solLinear[GammaL, DL, AL][t]]<0.001
\end{verbatim}
This fixes the value of \texttt{M} at the given $\textbf{k}$ and gives:
\begin{verbatim}
	{GammaN, DN, AN, lambdaN}
\end{verbatim}
\begin{table}[h]
	\centering
	\begin{tabular}{|c|c|c|c|c|}
		\hline
		$\text{k}$ & GammaN & DN & AN & lambdaN\\
		\hline
		\hline
		0.523599 & 0.00795374 & 0.0261371& 0.138968 &0.0000082\\
		\hline
		0.498666 &0.01020078  &0.02510015 & 0.085895&0.01000171\\
		\hline
		0.541654 &0.0100495 &0.0190012 & 0.0811839&-0.00291244\\
		\hline
		0.515015 & 0.00503164& 0.0361959&0.0540599& -0.0361143\\
		\hline
		0.546364 & 0.0092927& 0.022513&0.0540689& -0.0204979\\
		\hline
		0.537024 & 0.00493963&0.0356924 &0.0354816& 0.0980669\\
		\hline
		0.610018 & 0.00498291&0.0308737 &0.0330543&-0.0705018 \\
		\hline
		0.506708 & 0.00822755& 0.0233148& 0.104232 &0.0284448 \\
		\hline
	\end{tabular}
	\caption{Nonlinear fitting results.} 
	\label{tab}
\end{table}
%
Results are given in Table.~\ref{tab}. 
\vspace{2mm}

{
Let us  comment on the fitting method and its impact on the results. As mentioned before, in order to advance linear and nonlinear fits consistently, we choose to use a common } 
\verb|Method|  
{for both fits. Of the existing } 
\verb|Method|'s 
{in Mathematica, we found only one useful: }
\verb|NMinimize|. 
{For some other methods, such as }
\verb|Gradient| and \verb|Newton|, 
{we can only fit the data with numerical solutions to ``linear" equations. For the } 
\verb|LevenbergMarquardt| 
{method we cannot even find any fit to the linear equation.}
\begin{figure}[htbp]
	\centering
	\includegraphics[width=0.47\textwidth]{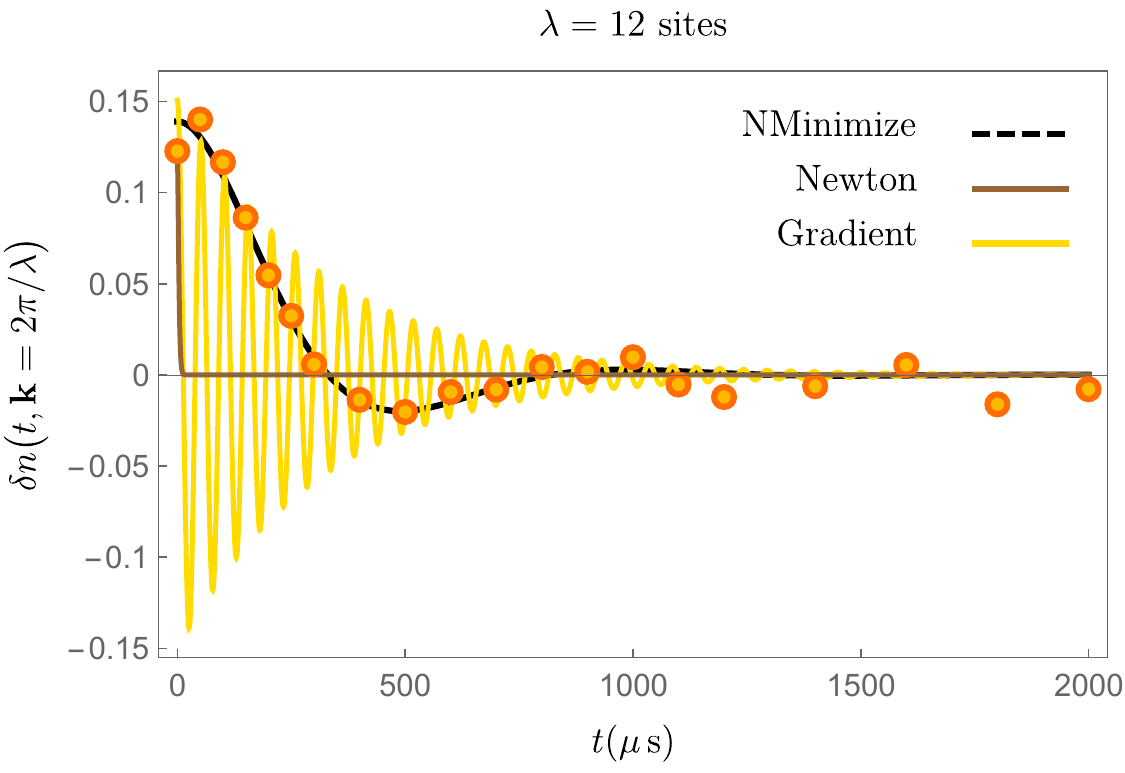}
	\caption{Comparing three different linear fitting methods to the data at a specific $\text{k}$ (from Ref. ~[21]). The black dashed curve corresponds to the fitting parameters given in the first row of Table ~I.	}
	\label{Fit}
\end{figure}
\vspace{2mm}

{In Fig.~\ref{Fit} we show the results of a linear fit to data with a specific value $\text{k}$ (or equivalent $\lambda$) found by three different methods. We see that while Mathematica is able to produce convergent results for all three cases. however, this is only }
\verb|NMinimize| 
{that can be considered a suitable method. It is worth noting that for some data sets from \cite{Brown:2018}, we find the results of the }
\verb|Newton| and \verb|Gradient| 
{to be divergent functions of time. But the results of } 
\verb|NMinimize| 
{method, given in Table.~\ref{table_near_horizon}, always converged and performed well for all eight sets of data from \cite{Brown:2018}.
\vspace{2mm}

With all the above points and observations, we are led to use the } 
\verb|NMinimize| 
{method. It should be noted that it is not surprising to find that only one of the above methods is suitable for our problem. In any fitting problem related to the solution of differential equations, the appropriate choice of fitting method depends largely on several factors:
\begin{itemize}
\item The fitted equation and whether the equation has an analytical solution.
\item The number of parameters in the equation plus the number of parameters in the boundary conditions.
\item The number of data points in any data set and their distribution.
\end{itemize}
Different fitting methods often do not give the same results for a specific problem.
}

\section{Application near a critical point}
{
Fluctuations in the UV-regulated theory of diffusion are also important near a critical point. Let us suppose a simple diffusive charge near the critical point. In the dynamic model H~\cite{hohenberg1977theory}, 
\begin{equation*}
	D\sim\xi^{-1}\,,\,\,\,\,\,\,\,\,\chi\sim \xi^2\,,\,\,\,\,\,\,\,\,\,\kappa\sim\xi \, ,
\end{equation*}
with $\xi$ being the correlation length. It turns out that $\tau\sim\xi^2$ 
\so{in agreement with the causality constraints}~\cite{Du:2021zqz}.
Near the critical point $\xi$ becomes large, so does the relaxation time $\tau$. As a result, the UV mode of our theory becomes a slow mode and then has to be included even in the long-wavelength limit~\cite{Stephanov:2017ghc}. 
On the other hand, we find that $\Sigma_3\sim\xi^{9/2}$, becoming large near the critical point. This shows the need to consider non-linearities discussed in this paper. \\
Within a QGP droplet passing by the conjectured QCD critical point, diffusion is coupled to bulk dynamics \cite{Du:2020bxp}. The 
\so{UV-regulated diffusion} 
model discussed in this paper then needs to be coupled with relativistic hydrodynamic fluctuations. 
The effect of interactions and self-interactions in non-linear MIS theory on the final proton multiplicity cumulants should then be investigated~\cite{Stephanov:2008qz}, and is relevant to the search for the critical point~\cite{Stephanov:1998dy}. 
}

%
